\documentclass[11pt]{JHEP3} 

\usepackage{epsfig,multicol,bbm}
\usepackage{amsmath}

\newcommand{\non}{\nonumber}
\newcommand{\del}{\partial}

\newcommand{\Vac}[1]{\bigg\langle{#1}\bigg\rangle}

\def\gtsim{\mathrel{\hbox{\raise0.2ex
\hbox{$>$}\kern-0.75em\raise-0.9ex\hbox{$\sim$}}}}
\def\ltsim{\mathrel{\hbox{\raise0.2ex
\hbox{$<$}\kern-0.75em\raise-0.9ex\hbox{$\sim$}}}}

\title{Electroweak phase transitions in the secluded $U(1)'$-extended MSSM}

\author{Cheng-Wei Chiang$^{1,2}$ and Eibun Senaha$^{1,3}$ \\
	$^1$Department of Physics and Center for Mathematics and Theoretical Physics, 
	National Central University, 300 Jhongda Rd., Jhongli, Taiwan 320, R.O.C.\\
	$^2$Institute of Physics, Academia Sinica, 128 Sec. 2, Academia Rd.,
	Nankang, Taipei, Taiwan 115, R.O.C.\\
	$^3$Physics Division, National Center for Theoretical Sciences, 
	101, Sec. 2 Kuang Fu Rd., Hsinchu, Taiwan 300, R.O.C. \\
	E-mail: \email{chengwei@phy.ncu.edu.tw, senaha@phys.cts.nthu.edu.tw}
	}

        \abstract{The electroweak phase transition (EWPT) in the secluded-$U(1)'$-extended MSSM (sMSSM) is studied.  Using the effective potential at zero and finite temperatures, we search for the non-MSSM-like EWPT in which the light stop mass is larger than the top quark mass.  Scanning the parameters relevant to the EWPT, the upper limits of the Higgs boson masses, which are consistent with the strong first order EWPT, are derived.  For the lightest $CP$-even and -odd Higgs bosons, we find $m_{H_1}\ltsim 160$ GeV and $m_{A_1}\ltsim 250$ GeV, respectively.  In the sMSSM, the tree-level $CP$ violation is possible by the complex soft supersymmetry breaking masses.  It is observed that such a $CP$-violating effect does not spoil the strong first order EWPT for the typical parameter sets.}


\begin{document} 



\section{Introduction}

To explain the baryon asymmetry of the Universe (BAU) is one of the challenging problems in particle physics and cosmology.  From the latest cosmological observations, the BAU is found to be~\cite{Amsler:2008zzb}.
\begin{eqnarray}
\frac{n_B}{n_\gamma}=(4.7-6.5)\times 10^{-10}\quad (95\%~{\rm C.L.}),
\end{eqnarray}
where $n_B$ is the difference between the number density of baryons and that of antibaryons, and $n_\gamma$ denotes the photon number density.  If there is an inflation in the early Universe, any primordial BAU would be washed out.  Therefore, the BAU must arise dynamically after the inflation.  In order to generate the BAU from a baryon-symmetric universe, it is required that~\cite{Sakharov:1967dj} (1) baryon number ($B$) violation, (2) $C$ and $CP$ violation and (3) departure from thermal equilibrium.  The last condition is not mandatory if the $CPT$ theorem does not hold.  In principle the standard model (SM) can satisfy these three conditions. However, it turns out that the $CP$-violating phase in the Cabibbo-Kobayashi-Maswaka matrix~\cite{CKM} is way too small to generate the observed BAU~\cite{ewbg_sm_cp} and a strong first order electroweak phase transition (EWPT) cannot be realized with the viable Higgs mass, $m_h>114.4$ GeV~\cite{Amsler:2008zzb}, rendering condition (3) infeasible~\cite{crossover}.  So far, many baryogenesis scenarios beyond the SM have been proposed.  Among them electroweak baryogenesis~\cite{ewbg} is an attractive idea since it exclusively relies on electroweak physics which is testable at colliders or cosmological observations in the near future.  The minimal supersymmetric standard model (MSSM), which is one of the well-motivated candidates for new physics at TeV scale, may have a window to a successful baryogenesis~\cite{ewbg-mssm,Carena:2008vj,Funakubo:2002yb}.  However, due to the strong experimental constraints the viable region seems to be quite limited.

A simple way to extend the MSSM is to introduce a gauge singlet ($S$) into the superpotential.  Several versions of the singlet extended MSSMs have been proposed: the Next-to MSSM (NMSSM)~\cite{NMSSM,Funakubo:2004ka}, the nearly MSSM (nMSSM) or the minimal non-MSSM(MNMSSM)~\cite{nMSSM}, the $U(1)'$-extended MSSM (UMSSM)~\cite{UMSSM,Cvetic:1997ky,UMSSM2} and the secluded $U(1)'$-extended MSSM (sMSSM/S-model)~\cite{Erler:2002pr,Han:2004yd,Kang:2004pp,Chiang:2008ud,Kang:2009rd} etc.  In this class of the models, there is no fundamental $\mu$ parameter in the superpotential, and it is generated after $S$ develops its vacuum expectation value (VEV), {\it i.e.}, $\mu_{\rm eff}=\lambda v_S$ where $\lambda$ denotes the dimensionless coupling and $v_S$ is the VEV of $S$.  It thus gives a solution to the so-called $\mu$ problem.

It is known that the Higgs singlet which couples to the Higgs doublets can play a role in strengthening the first order EWPT, and a light stop is not necessarily lighter than top quark in contrast to the MSSM case.  A variety of patterns of the EWPT are possible in the NMSSM~\cite{Funakubo:2005pu}.  Detailed studies of the EWPT in the nMSSM can be found in Refs.~\cite{Menon:2004wv,Huber:2006wf}.  For the analysis of the EWPT in the UMSSM, see, for example, Refs.~\cite{EWPT_UMSSM}.  The possibility of electroweak baryogenesis in the sMSSM is outlined in the letter paper~\cite{Kang:2004pp}, and recently its full paper has come out~\cite{Kang:2009rd}.  It is demonstrated that the electroweak baryogenesis is successful in this model.  Although a number of studies on the EWPT in the singlet-extended MSSM can be found in the literature, it is still not clear how heavy Higgs boson can be consistent with the strong first order EWPT in some models.  Such a mass limit would be indispensable for a test of EW baryogenesis scenario at colliders.

In this paper, we examine the sMSSM EWPT with/without $CP$ violation in the wider parameter space which has not been probed in Refs.~\cite{Kang:2004pp,Kang:2009rd}.  Since there are the singlet contributions in the Higgs potential, the stability of the Higgs potential is not manifest.  In our analysis, we require that the prescribed EW vacuum should be a global minimum at zero temperature.  This vacuum condition can make the allowed region quite limited~\cite{Chiang:2008ud}.  To investigate the properties of the EWPT, we use the one-loop effective potential at zero and nonzero temperatures taking the contributions from the $Z$ and $W$ bosons, the third generation of quarks and squarks into account.  Under the theoretical and experimental constraints, we exclusively search for the non-MSSM-like EWPT by scanning the relevant parameters and work out the upper limits of the Higgs boson masses which are consistent with the strong first order EWPT.

In the sMSSM, owing to the soft supersymmetry (SUSY) breaking mass terms, it is possible to realize both explicit and spontaneous $CP$ violation at the tree level.  The effect of such a $CP$-violating phase on the strength of the first order EWPT and the Higgs mass spectrum are investigated.

In general, there are ten order parameters if spontaneous $CP$ violation exists.  It is a non-trivial task to investigate such an EWPT thoroughly, and the numerical calculation is extremely time consuming.  We will focus exclusively on both the $CP$-conserving case and the explicit $CP$-violating case in which the number of order parameters is reduced to six.

The paper is organized as follows.  In Sec.~\ref{sec:Model}, we briefly explain the model and define the notations.  Section \ref{sec:Higgs_vac} is devoted to the tree-level analysis.  The approximate formulae of the Higgs spectrum are presented.  We discuss the vacuum condition, which turns out to be the severest theoretical constraint in this study.  In Sec.~\ref{sec:EWPT}, we argue the EWPT qualitatively in some detail.  To get some idea about the sMSSM EWPT, we outline the nMSSM EWPT by showing the analytical formulae.  In Sec.~\ref{sec:num_result}, the numerical results are presented.  Sec.~\ref{sec:conclusion} contains the conclusions and discussions.


\section{The model}\label{sec:Model}

The sMSSM is one of the singlet-extended MSSM models, which can be regarded as an effective theories of some unification theory such as string theory.  The symmetry of the model is $SU(3)_C\times SU(2)_{L}\times U(1)_{Y}\times U(1)'_{Q'}$, where the extra $U(1)$ is a remnant of the larger symmetry of the UV theory.  The Higgs sector comprises two Higgs doublets ($H_d, H_u$) and four Higgs singlets ($S, S_1, S_2, S_3$).  Among the singlets, the so-called secluded singlets ($S_i, i=1,2,3.$) play an essential role in ameliorating the severe experimental constraints on the $Z'$ boson~\cite{Erler:2002pr}.

It is desirable to have $U(1)'$ charges ($Q$'s) chosen to make the model
anomaly free. To this end, exotic chiral supermultiplets are
generally required~\cite{Cvetic:1997ky,Kang:2009rd,Erler:2000wu,Kang:2004bz}.
For our purpose, we assume that they are heavy enough not to affect the phenomenology at the electroweak scale.  Neither will we address the gauge coupling unification issue here as it requires the knowledge of full particle spectrum in the model.  Instead, we focus exclusively on the Higgs sector to discuss the EWPT.  Here, the so-called Model I sMSSM is considered, 
which can accommodate $CP$ violation at the tree level.  In Model I, the $U(1)'$ charges of the Higgs fields must satisfy
\begin{eqnarray}
Q_{H_d}+Q_{H_u}+Q_S=0,\quad
Q_S=-Q_{S_1}=-Q_{S_2}=\frac{1}{2}Q_{S_3}.\label{U1charge}
\end{eqnarray}
The superpotential ($W$) includes the following trilinear terms:
\begin{eqnarray}
W\ni-\epsilon_{ij}\lambda \widehat{S}\widehat{H}_d^i\widehat{H}_u^j
	-\lambda_S\widehat{S}_1\widehat{S}_2\widehat{S}_3 ~,
\end{eqnarray}
where $\widehat{H}_{d,u}$, $\widehat{S}, \widehat{S}_i, i=1,2,3,$ are the chiral superfields, $\lambda$ and $\lambda_S$ are the dimensionless couplings.  Due to the $U(1)'$ symmetry, the $S^n~(n\in \mathbb{Z})$ terms are forbidden.  Therefore, we are not bothered by the domain wall problem, which can be induced by the spontaneous breaking of the discrete symmetries.  
In general, the terms of the form $\widehat{S}\widehat{S}_1$, $\widehat{S}\widehat{S}_2$, $\widehat{S}^2_1\widehat{S}_3$ and $\widehat{S}^2_2\widehat{S}_3$
are allowed under the charge assignments (\ref{U1charge}).
However, we will not consider such quadratic terms since they may reintroduce the $\mu$ problem.
Neither will we consider the cubic terms for simplicity.
As we mentioned in the Introduction, once the Higgs singlet $S$ develops its VEV, the effective $\mu$ term is generated by $\mu_{\rm eff}=\lambda\langle S\rangle$.  Therefore, the scale of $\mu_{\rm eff}$ is not completely arbitrary but is fixed by the soft SUSY breaking parameters.

%
%
The Higgs potential at the tree level is given by
\begin{eqnarray}
V_0=V_F+V_D+V_{\rm soft},
\end{eqnarray}
with
\begin{eqnarray}
V_F&=&|\lambda|^2\big\{|\epsilon_{ij}\Phi_d^i\Phi_u^j|^2+|S|^2(\Phi_d^\dagger\Phi_d
	+\Phi_u^\dagger\Phi_u)\big\}
	+|\lambda_S|^2(|S_1S_2|^2+|S_2S_3|^2+|S_3S_1|^2),\\
V_D&=&\frac{g_2^2+g_1^2}{8}(\Phi_d^\dagger\Phi_d-\Phi_u^\dagger\Phi_u)^2 
	+\frac{g_2^2}{2}|\Phi_d^\dagger\Phi_u|^2\non\\
	&&+\frac{g'^2_1}{2}\Big(Q_{H_d}\Phi_d^\dagger\Phi_d
	+Q_{H_u}\Phi_u^\dagger\Phi_u+Q_S|S|^2+\sum_{i=1}^3Q_{S_i}|S_i|^2\Big)^2,\\
V_{\rm soft}&=&m_1^2\Phi_d^\dagger\Phi_d+m_2^2\Phi_u^\dagger\Phi_u 
	+m_S^2|S|^2+\sum_{i=1}^3m_{S_i}^2|S_i|^2\non\\
	&&-(\epsilon_{ij}\lambda A_{\lambda}S\Phi_d^i\Phi_u^j 
	+\lambda_SA_{\lambda_S}S_1S_2S_3
	+m_{SS_1}^2SS_1+m_{SS_2}^2SS_2+m_{S_1S_2}^2S_1^\dagger S_2+{\rm h.c.}),\non\\
\end{eqnarray}
where $g_2$, $g_1$ and $g'_1$ are the $SU(2)$, $U(1)$ and $U(1)'$ gauge couplings, respectively.  We will take $g'_1=\sqrt{5/3}g_1$ as motivated by the gauge unification in the simple GUTs. The soft SUSY breaking masses $m^2_{SS_1}$ and $m^2_{SS_2}$ are introduced to break two unwanted global $U(1)$ symmetries, and $m^2_{S_1S_2}$ is needed for explicit $CP$ violation (ECPV).

Here, we will follow the notation given in Ref.~\cite{Chiang:2008ud}.  The Higgs VEVs and their fluctuation fields are parameterized as
\begin{eqnarray}
\Phi_d&=&
e^{i\theta_1}\left(
\begin{array}{c}
\frac{1}{\sqrt{2}}(v_d+h_d+ia_d) \\
\phi_d^-
\end{array}
\right),\quad 
\Phi_u=
e^{i\theta_2}\left(
\begin{array}{c}
\phi_u^+\\
\frac{1}{\sqrt{2}}(v_u+h_u+ia_u) 
\end{array}
\right), \\
S&=&\frac{e^{i\theta_S}}{\sqrt{2}}(v_S+h_S+ia_S), \quad
S_i=\frac{e^{i\theta_{S_i}}}{\sqrt{2}}(v_{S_i}+h_{S_i}+ia_{S_i}),\quad i=1-3,
\end{eqnarray}
where $v=\sqrt{v^2_d+v^2_u}\simeq 246$ GeV at the vacuum.  The nonzero $\theta$'s can bring about spontaneous $CP$ violation (SCPV).  However, not all $\theta$'s are independent.  Due to the gauge invariance, the following four combinations are physical:
\begin{eqnarray}
\varphi_1=\theta_S+\theta_{S_1},\quad \varphi_2=\theta_S+\theta_{S_2},\quad
\varphi_3=\theta_S+\theta_1+\theta_2,\quad
\varphi_4=\theta_{S_1}+\theta_{S_2}+\theta_{S_3}.
\end{eqnarray}
The analysis of SCPV at zero temperature can be found in Ref.~\cite{Chiang:2008ud}.  To accommodate SCPV, the lightest Higgs boson mass should be less than about 125 GeV.  Although it is interesting to study SCPV at finite temperature, we will not pursue this possibility in this paper.  A comprehensive study of the SCPV at the finite temperature can be found in Ref.~\cite{Kang:2009rd}.

The tadpole conditions, which are defined by the first derivative of the Higgs potential
with respect to the Higgs fields, are given by
\begin{eqnarray}
\frac{1}{v_d}\Vac{\frac{\del V_0}{\del h_d}}
&=&m_{1}^{2}+\frac{g_{2}^{2}+g_{1}^{2}}{8}(v_{d}^{2}-v_{u}^{2})-R_\lambda\frac{v_uv_S}{v_d}
	+\frac{|\lambda|^2}{2}(v_u^2+v_S^2)+\frac{g'^2_1}{2}Q_{H_d}\Delta=0,\label{tad_hd}
\nonumber \\\\
\frac{1}{v_u}\Vac{\frac{\del V_0}{\del h_u}}
&=&m_{2}^{2}-\frac{g_{2}^{2}+g_{1}^{2}}{8}(v_{d}^{2}-v_{u}^{2})-R_\lambda\frac{v_dv_S}{v_u}
	+\frac{|\lambda|^2}{2}(v_d^2+v_S^2)+\frac{g'^2_1}{2}Q_{H_u}\Delta=0,\label{tad_hu}
\nonumber \\\\
\frac{1}{v_S}\Vac{\frac{\del V_0}{\del h_S}}
&=&m_S^2-(R_1v_{S_1}+R_2v_{S_2}+R_\lambda v_dv_u)\frac{1}{v_S}
	+\frac{|\lambda|^2}{2}(v_d^2+v_u^2)+\frac{g'^2_1}{2}Q_{S}\Delta=0,\label{tad_hS}
\nonumber \\\\
\frac{1}{v_{S_1}}\Vac{\frac{\del V_0}{\del h_{S_1}}}
&=&m_{S_1}^2-(R_1v_{S}+R_{12}v_{S_2}+R_{\lambda_S}v_{S_2}v_{S_3})
	\frac{1}{v_{S_1}}\non\\
&&\hspace{1cm}
	+\frac{|\lambda_S|^2}{2}(v_{S_2}^2+v_{S_3}^2)+\frac{g'^2_1}{2}Q_{S_1}\Delta=0,
	\label{tad_hS1}\\
\frac{1}{v_{S_2}}\Vac{\frac{\del V_0}{\del h_{S_2}}}
&=&m_{S_2}^2-(R_2v_{S}+R_{12}v_{S_1}+R_{\lambda_S}v_{S_1}v_{S_3})
	\frac{1}{v_{S_2}}\non\\
&&\hspace{1cm}
	+\frac{|\lambda_S|^2}{2}(v_{S_1}^2+v_{S_3}^2)+\frac{g'^2_1}{2}Q_{S_2}\Delta=0,
	\label{tad_hS2}\\
\frac{1}{v_{S_3}}\Vac{\frac{\del V_0}{\del h_{S_3}}}
&=&m_{S_3}^2-R_{\lambda_S}\frac{v_{S_1}v_{S_2}}{v_{S_3}}
	+\frac{|\lambda_S|^2}{2}(v_{S_1}^2+v_{S_2}^2)
	+\frac{g'^2_1}{2}Q_{S_3}\Delta=0,\label{tad_hS3}\\
\frac{1}{v_u}\Vac{\frac{\del V_0}{\del a_d}}
&=&\frac{1}{v_d}\Vac{\frac{\del V_0}{\del a_u}}=I_\lambda v_S=0,\label{tad_ad}\\
\Vac{\frac{\del V_0}{\del a_S}}
&=&I_1v_{S_1}+I_2v_{S_2}+I_\lambda v_dv_u=0,\label{tad_aS}\\
\Vac{\frac{\del V_0}{\del a_{S_1}}}
&=&I_1v_{S}-I_{12}v_{S_2}+I_{\lambda_S}v_{S_2}v_{S_3}=0,\label{tad_aS1}\\
\Vac{\frac{\del V_0}{\del a_{S_2}}}
&=&I_2v_{S}+I_{12}v_{S_1}+I_{\lambda_S}v_{S_1}v_{S_3}=0,\label{tad_aS2}\\
\Vac{\frac{\del V_0}{\del a_{S_3}}}&=&I_{\lambda_S}v_{S_1}v_{S_2}=0,\label{tad_aS3}
\end{eqnarray}
where
\begin{eqnarray}
\Delta&=&Q_{H_d}v_d^2+Q_{H_u}v_u^2+Q_{S}v_S^2+\sum_{i=1}^3Q_{S_i}v_{S_i}^2,\\
R_i&=&{\rm Re}(m^2_{SS_i}),\quad
I_i={\rm Im}(m^2_{SS_i}),\quad i=1,2, \\
R_{12}&=&{\rm Re}(m^2_{S_1S_2}),\quad 
I_{12}={\rm Im}(m^2_{S_1S_2}),\\
R_\lambda&=&\frac{{\rm Re}(\lambda A_\lambda)}{\sqrt{2}},\quad 
I_\lambda=\frac{{\rm Im}(\lambda A_\lambda)}{\sqrt{2}},\\
R_{\lambda_S}&=&\frac{{\rm Re}(\lambda_S A_{\lambda_S})}{\sqrt{2}},\quad 
I_{\lambda_S}=\frac{{\rm Im}(\lambda_S A_{\lambda_S})}{\sqrt{2}},
\end{eqnarray}
and $\langle X\rangle$ is defined such that $X$ is evaluated at the vacuum.  In our analysis, the soft SUSY breaking masses ($m^2_1$, $m^2_2$, $m^2_S$, $m^2_{S_i}, i=1,2,3.$) are determined via the six tadpole conditions (\ref{tad_hd})-(\ref{tad_hS3}).  Here, all the Higgs VEVs are regarded as the input parameters and assumed to be nonzero.

From the tadpole conditions with respect to the $CP$-odd Higgs fields, it follows that
\begin{eqnarray}
I_\lambda=I_{\lambda_S}=0,\quad I_1=I_{12}\frac{v_{S_2}}{v_S},\quad
I_2=-I_{12}\frac{v_{S_1}}{v_S}.
\label{CPV_vac}
\end{eqnarray}
Therefore, there is only one physical $CP$-violating phase at the tree level, and we take ${\rm Arg}(m^2_{S_1S_2})\equiv\theta_{S_1S_2}$ as an input.  After including the one-loop contributions, the relations (\ref{CPV_vac}) will be modified, and will be discussed in subsection~\ref{subsec:CPV}.


\section{Higgs mass spectrum and vacuum conditions}\label{sec:Higgs_vac}

In this paper, we consider the one-loop corrections from the $Z$ and $W$ bosons, 
the third generation of quarks ($t, b$), and squarks ($\tilde{t}_{1,2}, \tilde{b}_{1,2}$).  The one-loop effective potential at zero temperature takes the form~\cite{Coleman:1973jx}
\begin{eqnarray}
V_1(\Phi_d,\Phi_u,S,S_{1,2,3})
&=&\sum_Ac_A\frac{\bar{m}^4_A}{64\pi^2}\left(\ln\frac{\bar{m}^2_A}{M^2}-\frac{3}{2}\right),
\label{Veff_1loop}
\end{eqnarray}
which is regularized in the $\overline{\rm DR}$ scheme, $\bar{m}_A$ is a field dependent mass, and $M$ is the renormalization scale determined by the condition $\langle V_1\rangle=0$.  The statistical factor of each particle is given respectively by $c_Z=3,~c_W=6,~c_t=c_b=-4N_C,$ and $c_{\tilde{t}_{1,2}}=c_{\tilde{b}_{1,2}}=2N_C$, where $N_C$ is the color factor.

In principle, the $Z$ boson can mix with the $Z'$ boson, and their mass matrix becomes 2-by-2.  However, the mass of $Z'$ boson and the magnitude of the mixing angle are strongly constrained by experiments.  A recent analysis of the constraints on the $Z'$ boson mass and the mixing angle can be found in Ref.~\cite{Zprime_search}.
As the mixing angle is constrained to be less than $10^{-3}$~\cite{Zprime_search}, we simply consider the case of no $Z$-$Z'$ mixing, {\it i.e.}, $\tan\beta = \sqrt{Q_{H_d}/Q_{H_u}}$.  In addition, since the $\tan\beta \simeq \mathcal{O}(1)$ is generic in the sMSSM~\cite{Erler:2002pr,Han:2004yd,Chiang:2008ud}, we will present only the case of $\tan\beta=1$.  In this case, the secluded sector does not contribute to Eq.~(\ref{Veff_1loop}).


\subsection{Higgs mass spectrum}\label{subsec:Higgs_spectrum}

Due to the additional contributions coming from the $F$-term and $D$-term of $U(1)'$, the mass bound on the lightest Higgs boson is significantly relaxed compared to the MSSM.  At the tree level, it is found that
\begin{eqnarray}
  m^2_{H_1}&\le& m^2_Z\cos^22\beta+\frac{|\lambda|^2}{2}v^2\sin^22\beta
	+g'^2_1v^2(Q_{H_d}\cos^2\beta+Q_{H_u}\sin^2\beta)^2 \non\\
	&=&(0~{\rm GeV})^2+(139~{\rm GeV})^2+(111~{\rm GeV})^2\simeq(178~{\rm GeV})^2, 
\end{eqnarray}
where we have taken $\tan\beta=1,~\lambda=0.8,~Q_{H_d}=Q_{H_u}=1$ in the second line.  Thus, large radiative corrections are not necessarily required for avoiding the LEP exclusion mass limits.  However, we should note that because of the mixing terms between the doublets and singlets, $m^2_{H_1}$ can become smaller and even negative.  For the above parameter set, we get
\begin{eqnarray}
  m^2_{H_1} = \frac{1}{2}\left[
	m^2_S+|\lambda|^2v^2+6g'^2_1v^2_S
	-\sqrt{\Big\{m^2_S+2g'^2_1(3v^2_S-v^2)\Big\}^2
	+4v^2\Big\{R_\lambda-(|\lambda|^2-2g'^2_1)v_S\Big\}^2 }
	\right],\label{mH1_app}\non\\
\end{eqnarray}
where the mixing terms coming from the secluded sector are neglected, 
and $CP$ is assumed to be conserved for simplicity.  The parameter $m^2_S$ appearing in Eq.~(\ref{mH1_app}) is given by the tadpole condition for
$h_S$, {\it i.e.}, Eq.~(\ref{tad_hS}).  Therefore, $m^2_{H_1}$ can become negative in the large $R_\lambda$ limit.  It should be noted that $R_\lambda$ can be re-expressed in terms of $m_{H^\pm}$.  At the tree level,
\begin{eqnarray}
  m_{H^\pm}^2=\frac{1}{\sin\beta\cos\beta}\Vac{\frac{\del^2 V_0}{\del\phi_d^+\del\phi_u^-}}
	=m_W^2+\frac{2R_\lambda}{\sin2\beta}v_S-\frac{|\lambda|^2}{2}v^2.\label{mch}
\end{eqnarray}
Thus, $m_{H_1}$ can be unphysical in the large $m_{H^\pm}$ for a moderate value of $v_S$.  The one-loop formula of $m^2_{H^\pm}$ is explicitly given in Ref.~\cite{Chiang:2008ud}.  As is done there, we take $m_{H^\pm}$ as an input in place of $|A_\lambda|$.

To obtain more precise values in the Higgs mass spectrum, it is necessary to incorporate the mixing terms between $(\Phi_d, \Phi_u, S)$ and $(S_1, S_2, S_3)$ sectors, and also the one-loop contributions (denoted by $\Delta m^2_H$).  It is well-known that the dominant term of $\Delta m^2_H$ comes from the top/stop loops:
\begin{eqnarray}
  \Delta m^2_H = \frac{N_C}{16\pi^2}\frac{8}{v^2}
	m^4_t\ln\frac{m_{\tilde{t}_1}m_{\tilde{t}_2}}{m^2_t}.
\end{eqnarray}
The explicit formulae of the mass matrix at the tree level are presented in Refs.~\cite{Erler:2002pr,Chiang:2008ud}.  For the one-loop expression, which is the same as the that of the NMSSM, see, for example, Ref.~\cite{Funakubo:2004ka}.  After the $SU(2)$ Nambu-Goldstone bosons are rotated away, the mass matrix of the neutral Higgs bosons can be reduced to an 11-by-11 form. We diagonalize it numerically to obtain $m_{H_i}$, where the subscript $i$ is labeled in the ascending order of mass.

Here, we comment on the possible range of $m_{H^\pm}$.  It is known that $m_{H^\pm}$ is constrained from above by the vacuum condition.  Namely, if we require that the energy level of the electroweak vacuum $v=\sqrt{v^2_d+v^2_u}\simeq 246$ GeV should be lower than the origin at which the energy level is normalized to zero, $m_{H^\pm}$ cannot exceed some critical value.  For the typical parameter sets, the maximal value of $m_{H^\pm}$ is found to be $\mathcal{O}(1-10)$ TeV~\cite{Chiang:2008ud}.  Actually, we can get a stronger constraint since the symmetric vacuum where $v_d=v_u=0$ is not necessarily located at the origin.  We will discuss the vacuum condition in the next subsection.


\subsection{Energy levels of the vacua}\label{Elevel_vacua}

Before moving on to the analysis of the EWPT, we consider the structure of the zero-temperature effective potential in some detail.  Although we restrict ourselves to the tree-level analysis for simplicity, it suffices to know the qualitative features of the Higgs potential in the sMSSM.

The effective potential at the tree level takes the form
\begin{eqnarray}
V_0&=&\frac{1}{2}m_1^2v_d^2+\frac{1}{2}m_2^2v_u^2+\frac{1}{2}m_S^2v_S^2
	+\sum_i\frac{1}{2}m_{S_i}^2v_{S_i}^2
	-R_1v_Sv_{S_1}-R_2v_Sv_{S_2}-R_{12}v_{S_1}v_{S_2}\non\\
&&-R_\lambda v_dv_uv_S-R_{\lambda_S}v_{S_1}v_{S_2}v_{S_3}
	+\frac{g_2^2+g_1^2}{32}(v_d^2-v_u^2)^2
	+\frac{|\lambda|^2}{4}(v_d^2v_u^2+v_d^2v_S^2+v_u^2v_S^2) \non\\
&&	+\frac{|\lambda_S|^2}{4}(v_{S_1}^2v_{S_2}^2+v_{S_2}^2v_{S_3}^2+v_{S_3}^2v_{S_1}^2)
	+\frac{g'^2_1}{8}\Delta^2.\label{V0}
\end{eqnarray}

The EW vacuum defined by the tadpole conditions (in what follows we refer it as the prescribed EW vacuum) is not always the global minimum.  In order to see this, we compare the energy levels of the EW vacuum and the symmetric vacuum.  The energy levels of the two vacua are given respectively by
\begin{eqnarray}
  \langle V_0\rangle_{\rm vac}
  &=&\frac{1}{2}R_\lambda v_dv_uv_S+\frac{1}{2}R_{\lambda_S}v_{S_1}v_{S_2}v_{S_3}
	-\frac{g_2^2+g_1^2}{32}(v_d^2-v_u^2)^2 \non\\
&&-\frac{|\lambda|^2}{4}(v_d^2v_u^2+v_d^2v_S^2+v_u^2v_S^2)
	-\frac{|\lambda_S|^2}{4}(v_{S_1}^2v_{S_2}^2+v_{S_2}^2v_{S_3}^2+v_{S_3}^2v_{S_1}^2)
	-\frac{g'^2_1}{8}\Delta^2, \non\\\\
\langle V^{({\rm sym})}_0\rangle_{\rm vac}
&=&\langle V_0(v_{d,u}=0)\rangle_{\rm vac} \non\\
&=&\frac{1}{2}R_{\lambda_S}\bar{v}_{S_1}\bar{v}_{S_2}\bar{v}_{S_3}
	-\frac{|\lambda_S|^2}{4}(\bar{v}_{S_1}^2\bar{v}_{S_2}^2
	+\bar{v}_{S_2}^2\bar{v}_{S_3}^2+\bar{v}_{S_3}^2\bar{v}_{S_1}^2)
	-\frac{g'^2_1}{8}\bar{\Delta}^2.
\end{eqnarray}
where $\bar{v}$'s are determined by the tadpole conditions using the potential $V^{({\rm sym})}_0$, and $\bar{\Delta}$ is given by $\Delta(v=\bar{v})$.  The energy level of the symmetric vacuum can in principle become higher than the origin in the limit of large positive $R_{\lambda_S}$.

The difference between the energy levels of the two vacua is 
\begin{eqnarray}
\Delta\langle V_0\rangle_{\rm vac}&=&
\langle V^{({\rm sym})}_0\rangle_{\rm vac}-\langle V_0\rangle_{\rm vac} \non\\
&=&-\frac{1}{2}R_\lambda v_dv_uv_S
	+\frac{1}{2}R_{\lambda_S}(\bar{v}_{S_1}\bar{v}_{S_2}\bar{v}_{S_3}
	-v_{S_1}v_{S_2}v_{S_3}) \non\\
&&	+\frac{g^2_2+g^2_1}{32}(v^2_d-v^2_u)^2
	+\frac{|\lambda|^2}{4}(v^2_dv^2_u+v^2_dv^2_S+v^2_uv^2_S) \non\\
&&	-\frac{|\lambda_S|^2}{4}
	\Big[\bar{v}_{S_1}^2\bar{v}_{S_2}^2+\bar{v}_{S_2}^2\bar{v}_{S_3}^2
		+\bar{v}_{S_3}^2\bar{v}_{S_1}^2
		-v_{S_1}^2v_{S_2}^2-v_{S_2}^2v_{S_3}^2-v_{S_3}^2v_{S_1}^2
		\Big] \non\\
&&	-\frac{g'^2_1}{8}(\bar{\Delta}^2-\Delta^2).
\label{del_Vmin}
\end{eqnarray}
For a relatively large $R_\lambda$, $\Delta\langle V_0\rangle_{\rm vac}<0$ can happen; namely, the energy level of the symmetric vacuum becomes lower than that of the broken one.  Although such an EW vacuum could have a long lifetime that is larger than the age of the Universe and thus look viable, it is a highly nontrivial question whether the EW symmetry can be restored as the temperature increases. We thus exclude it from the investigation\footnote{In the viable MSSM baryogenesis scenario, the EW vacuum is metastable and sufficiently long-lived, and the charge-color-breaking vacuum is the global minimum.  On the other hand, the energy level of the EW symmetric vacuum is higher than that of the broken one as in the usual scenario.  In such a case, the successful EW symmetry restoration is possible~\cite{Carena:2008vj}.}.  As mentioned in subsection \ref{subsec:Higgs_spectrum}, a large $R_\lambda$ is nothing but a large $m_{H^\pm}$.  Therefore, the maximally allowed value of $m_{H^\pm}$ can be derived from this vacuum condition.

\begin{figure}[t]
\begin{center}
\includegraphics[width=7cm]{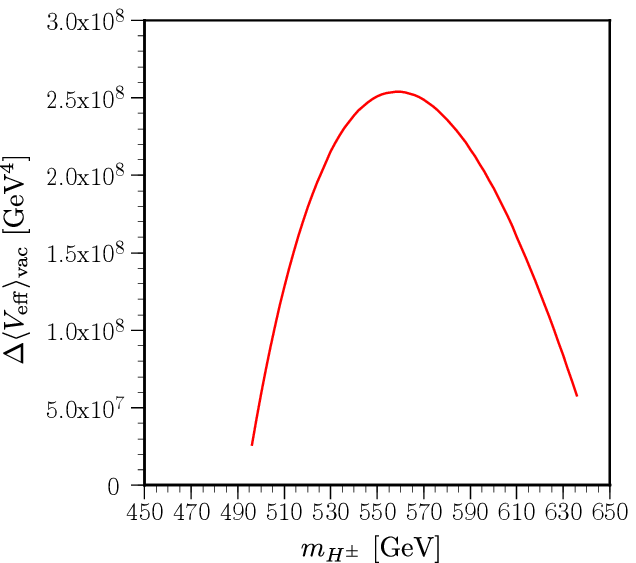}
\hspace{1cm}
\includegraphics[width=6.5cm]{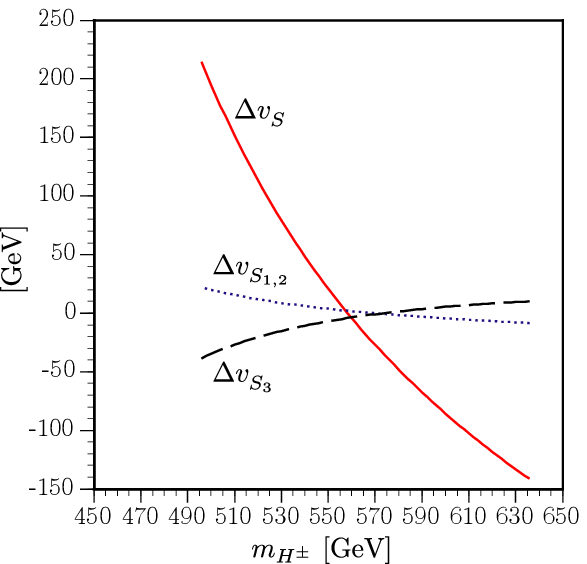}
\caption{Left: $\Delta\langle V_{\rm eff}\rangle_{\rm vac} =\langle V^{({\rm sym})}_{\rm eff}\rangle_{\rm vac} -\langle V_{\rm eff}\rangle_{\rm vac}$ as function of $m_{H^\pm}$.  Right: $\Delta v_S=\bar{v}_S-v_S$ and $\Delta v_{S_i}=\bar{v}_{S_i}-v_{S_i}$, $i=1,2,3$, as a function of $m_{H^\pm}$.  We take $\tan\beta=1$, $\lambda=0.8$, $\lambda_S=0.3$, $v_S=500$ GeV, $ v_{S_1}=v_{S_2}=v_{S_3}=1200$ GeV, $m^2_{SS_1}=m^2_{SS_2}=(50~{\rm GeV})^2$, $m^2_{S_1S_2}=(200~{\rm GeV})^2$, $A_{\lambda_S}=A_\lambda$.}
\label{fig:Veff_tree}
\end{center}
\end{figure}

The signs of the second, fifth and last terms of Eq.~\ref{del_Vmin} depend on the magnitudes of $\bar{v}$'s.  In Fig.~\ref{fig:Veff_tree}, we plot various energy differences numerically.  The left panel shows $\Delta\langle V_{\rm eff} \rangle_{\rm vac}$, where the one-loop corrections, $V_{1}$ in Eq.~(\ref{Veff_1loop}), have been included in the numerical calculation\footnote{The behaviors of $\Delta\langle V_{\rm eff}\rangle_{\rm vac}$ and $\Delta\langle V_0\rangle_{\rm vac}$ are almost the same.}.  As an illustration, we take $\tan\beta=1$, $\lambda=0.8$, $\lambda_S=0.3$, $v_S=500$ GeV, $ v_{S_1}=v_{S_2}=v_{S_3}=1200$ GeV, $m^2_{SS_1}=m^2_{SS_2}=(50~{\rm GeV})^2$, $m^2_{S_1S_2}=(200~{\rm GeV})^2$, and $A_{\lambda_S}=A_\lambda$.  We call this parameter set Case 1, which we will discuss in greater detail in Sec.~\ref{sec:num_result}.  In this case, a stable EW vacuum ($\Delta\langle V_{\rm eff}\rangle_{\rm vac}>0$) exists for 496 GeV $\ltsim m_{H^\pm} \ltsim 636$ GeV.  As $m_{H^\pm}$ decreases, the fifth term with a negative coefficient becomes dominant and then eventually results in $\Delta\langle V_{\rm eff}\rangle_{\rm vac}<0$.  On the other hand, as $m_{H^\pm}$ increases, $\Delta\langle V_{\rm eff}\rangle_{\rm vac}<0$ happens due to the contribution of the $R_\lambda$ term with a negative coefficient as indicated by Eq.~(\ref{del_Vmin}).

In the right panel of Fig.~\ref{fig:Veff_tree}, we plot $\Delta v_S$ and $\Delta v_{S_i}$, $i=1,2,3$, where $\Delta v_S=\bar{v}_S-v_S$ and $\Delta v_{S_i}=\bar{v}_{S_i}-v_{S_i}$.  In the region where $\Delta\langle V_{\rm eff}\rangle_{\rm vac}$ is small, $|\Delta v_S|$ and $|\Delta v_{S_i}|$ become large.  As noticed in Ref.~\cite{Funakubo:2005pu}, the locations of the symmetric and broken vacua at {\it zero temperature} may yield some information about the EWPT. 
Namely, the sizable $|\Delta\bar{v}_S|$ indicates that the singlet Higgs may be
involved in realizing the non-MSSM-like EWPT.
In such a case, since $\Delta\langle V_{\rm eff}\rangle_{\rm vac}$ is small, the EW symmetry is expected to be restored at a relatively low temperature. 

\section{Electroweak phase transition}\label{sec:EWPT}
%
%
Here, we give the necessary ingredients for calculating the EWPT and describe the EWPT qualitatively.  The one-loop effective potential at finite temperature takes the form
\begin{eqnarray}
V_1(\Phi_d,\Phi_u,S,S_{1,2,3};T)&=&
\frac{T^4}{2\pi^2}\sum_Ac_AI_{B,F}\left(\frac{\bar{m}^2_A}{T^2}\right),
\end{eqnarray}
with
\begin{eqnarray}
I_{B, F}(a^2)&=&\int_0^\infty dx~x^2\ln\left(1\mp e^{-\sqrt{x^2+a^2}}\right),\label{IBF}
\end{eqnarray}
where the subscripts of $I_{B,F}(a^2)$ denote boson (fermion).  In order to reduce the computation time for the numerical integration in $I_{B,F}(a^2)$, we will use the fitting functions employed in Ref.~\cite{Funakubo:2009eg} instead of Eq.~(\ref{IBF}).  More explicitly,
\begin{eqnarray}
\tilde{I}_{B,F}(a^2)=e^{-a}\sum^N_{n=0}c^{b,f}_na^n,
\label{V1_fit}
\end{eqnarray}
are used, where $c^{b,f}_n$ are determined by the least square method.  For $N=40$, $|I_{B,F}(a^2)-\tilde{I}_{B,F}(a^2)|<10^{-6}$ for any $a$, which suffices in our investigation.  In the sMSSM, the structure of the tree-level potential is expected to be more important than the higher-order corrections unless a right-handed stop is lighter than top quark, which is the successful scenario of the MSSM EW baryogenesis.  In the following, we exclusively explore a non-MSSM-like EWPT, {\it i.e.}, the heavy stop case.  Hence, we will not include the two-loop contributions, and neither will we perform the ring-improvements in the effective potential for simplicity.

For the EW baryogenesis to work, the sphaleron process, which is active in the 
symmetric phase, must be decoupled when the EWPT completes.  In other words, the sphaleron rate in the broken phase should be less than the Hubble parameter at that moment.  Conventionally, the sphaleron decoupling condition is cast into the form
\begin{eqnarray} 
\frac{\rho^{}_E}{T_E}>\zeta,
\label{sph_dec}
\end{eqnarray}
where $T_E$ is the temperature at which the EWPT ends,
$\rho_E$ is defined by the vacuum expectation values of the two Higgs doublets at $T_E$, i.e.,
$\rho_E=\sqrt{\rho^2_d(T_E)+\rho^2_u(T_E)}$,
and $\zeta$ is an $\mathcal{O}(1)$ parameter which depends on the profile of the sphaleron, etc.  
It is, however, a non-trivial task to evaluate $T_E$ explicitly since a full knowledge of bubble dynamics is required.  We thus use the condition $\rho_C/T_C>\zeta$ instead of Eq.~(\ref{sph_dec}), assuming that the supercooling is not too large, where $T_C$ is defined by the temperature at which the effective potential has two degenerate minima and $\rho^{}_C$ is the VEV at $T_C$.  To know the value of $\zeta$ within $\mathcal{O}(10\%)$ accuracy, the sphaleron energy and zero-mode factors of the fluctuations around the sphaleron must be evaluated using the finite temperature effective potential.  According to a recent study of the sphaleron decoupling condition with such an accuracy, $\zeta\simeq 1.4$ in the MSSM~\cite{Funakubo:2009eg}.  In the NMSSM, the calculation of the sphaleron energy based on the tree-level potential has been done, and the sphaleron energy is found to be more or less the same as in the MSSM for a large portion of its parameter space~\cite{Funakubo:2005bu}.  Although a similar condition is expected for the sMSSM as well, we here adopt the rough criterion $\zeta=1$ for simplicity.

\subsection{Patterns of the EWPT}\label{subsec:pattern_EWPT}

As mentioned in Sec.~\ref{sec:Model}, we will not consider the case of SCPV, 
which reduces the number of order parameters relevant to the EWPT to six. 
Now let us introduce
\begin{eqnarray}
\boldsymbol{\rho} = (\rho_d, \rho_u, \rho_S, \rho_{S_1}, \rho_{S_2}, \rho_{S_3})
\end{eqnarray}
such that $\boldsymbol{\rho}(T=0)=\boldsymbol{v}\equiv({v_d, v_u, v_S, v_{S_1}, v_{S_2}, v_{S_3}})$.

As seen from Eq.~(\ref{V0}), the effective potential in the $\rho_S$ direction has the form
\begin{eqnarray}
V(\rho_S)&\ni& -(R_1\rho_{S_1}+R_2\rho_{S_2})\rho_S
	+\frac{1}{2}m^2_S\rho^2_S+\frac{g'^2_1Q^2_S}{8}\rho^4_S
	+\frac{g'^2_1}{4}Q_S\Big(\sum_iQ_{S_i}\rho^2_{S_i}\Big)\rho^2_S.
\end{eqnarray}
Since the linear term in $\rho_S$ can exist in principle, the point $(\rho_d, \rho_u, \rho_S)=(0,0,0)$ is not necessarily a local minimum.  The coefficient of $\rho_S$ term can vanish only when $\rho_{S_1}=\rho_{S_2}=0$\footnote{In the parameter space explored in this paper, 
${\rm sgn}(R_{1,2})=+1$ must be taken in order
to be consistent with the positivity of the Higgs squared mass.}.  Therefore, we may categorize the EWPT into the following two types:
\begin{enumerate}
\item[]
Type A: $(0,0,\bar{\rho}_S,\bar{\rho}_{S_1},\bar{\rho}_{S_2},\bar{\rho}_{S_3})~\to~
(\rho_d,\rho_u,\rho_S,\rho_{S_1},\rho_{S_2},\rho_{S_3})$,
\item[]
Type B: $(0,0,0,0,0,\bar{\rho}_{S_3})~\to~(\rho_d,\rho_u,\rho_S,\rho_{S_1},\rho_{S_2},\rho_{S_3})$,
\end{enumerate}
where the barred quantities denote the corresponding VEV's in the symmetric phase.  The phase transition pattern of the $U(1)'$ symmetry may be even more diverse, and a detailed analysis of it is beyond the scope of the paper.  In what follows, we will concentrate on the EWPT of Type A and B.
%
\begin{figure}[t]
\centerline{\includegraphics[width=5cm]{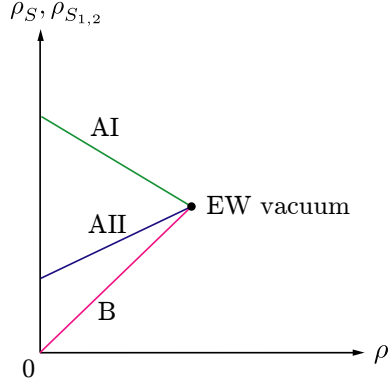}}
\caption{The patterns of the EWPT.}
\label{patterns_EWPT}
\end{figure}
%
We call it a Type AI EWPT if $\Delta v_S>0$ and a Type AII EWPT if $\Delta v_S<0$.
These types of EWPT are pictorially shown in Fig.~\ref{patterns_EWPT}
\footnote{Of course there is additional axis of $\rho_{S_3}$ which is not shown here.}.  In order to see the qualitative features of the EWPT, we consider a rather simplified case.  In the case where the temperature is high compared to the mass of the particle in the loop, $I_{B,F}(a^2)$ in Eq.~(\ref{IBF}) can be expanded in powers of $a^2=m^2/T^2$ as~\cite{Dolan:1973qd}
\begin{eqnarray}
  I_B(a^2)&=&-\frac{\pi^4}{45}+\frac{\pi^2}{12}a^2-\frac{\pi}{6}(a^2)^{3/2}
	-\frac{a^4}{32}\bigg(\ln\frac{a^2}{\alpha_B}-\frac{3}{2}\bigg)
	+\frac{\zeta(3)}{384\pi^2}a^6-\cdots, \label{IB_HTE}\\
I_F(a^2)&=&\frac{7\pi^4}{360}-\frac{\pi^2}{24}a^2
	-\frac{a^4}{32}\bigg(\ln\frac{a^2}{\alpha_F}
	-\frac{3}{2}\bigg)+\frac{7\zeta(3)}{384\pi^2}a^6-\cdots,
\end{eqnarray}
where
\begin{equation}
\ln\alpha_B=2\ln4\pi-2\gamma_E\simeq3.91,\qquad
\ln\alpha_F=2\ln\pi-2\gamma_E\simeq1.14,
\end{equation}
and $\gamma_E(\simeq 0.577)$ is the Euler constant and $\zeta(3)(\simeq1.202)$
is a Riemann zeta function value.  The relative errors of $I_{B,F}(a^2)$ are less than 5\% if $a\ltsim2.3$ for bosons and $a\ltsim1.7$ for fermions.

In the multi-Higgs-doublet models, the $(a^2)^{3/2}$ term with a negative coefficient in Eq.~(\ref{IB_HTE}) is crucial for the first order EWPT, as it gives rise to the potential barrier between the two degenerate minima.  Such a cubic term originates from the zero frequency modes in the bosonic thermal loop.

In the other limit, where the temperature is low compared to the mass, $I_{B,F}(a^2)$ can be expressed as~\cite{Anderson:1991zb}
\begin{eqnarray}
I_{B,F}(a^2)\simeq\mp a^2K_2(a)\simeq\mp \sqrt{\frac{\pi}{2}}a^{3/2}e^{-a}
	\left[1+\frac{15}{8a}+\cdots \right],
\end{eqnarray}
where $K_2(a)$ is the modified Bessel function.  The relative errors of $I_{B,F}(a^2)$ are less than 5\% if $a\gtsim 2.6$ for bosons and $a\gtsim 2.3$ for fermions.  For the EWPT, the typical critical temperature is $\mathcal{O}(100)$ GeV.  Therefore, thermal effects of the squarks considered herein are exponentially suppressed, and do not play a significant role in realizing the first order EWPT.
%
%
\subsection{Type A}\label{subsec:typeA}

We first begin with a simplified case.  Suppose that $g'_1=0$ and the VEVs of the secluded Higgs singlets are much larger than the others, which may correspond to the nMSSM-like limit.  Although the EWPT in the nMSSM has already been studied in Refs.~\cite{Menon:2004wv,Huber:2006wf}, we here give a brief review of it to get a qualitative feature of the EWPT in the sMSSM.  Since the nMSSM does not possess the cubic self-interacting term of the singlet Higgs boson in the Higgs potential, the analysis of the EWPT becomes much simpler than that of any other singlet-extended MSSMs.

Let us consider the EWPT in the subspace $\boldsymbol{\rho}=(\rho_d, \rho_u, \rho_S)$ assuming $\rho_{S_i}=v_{S_i}$.  Here we consider the tree-level potential and the dominant temperature-dependent contributions, which are proportional to $T^2$.  In the following, we show how the first order EWPT is realized without relying on the cubic term coming from the bosonic thermal loop as mentioned above.

The effective potential here is reduced to
\begin{eqnarray}
V(\boldsymbol{\rho}, T)= \frac{1}{2}M^2(T)\rho^2+\frac{1}{2}m^2_S\rho^2_S
	-(c+\tilde{R}_\lambda\rho^2)\rho_S+\frac{|\lambda|^2}{4}\rho^2\rho^2_S
	+\frac{\tilde{\lambda}^2}{4}\rho^4,\label{Veff_caseI}
\end{eqnarray}
where
\begin{eqnarray}
M^2(T) &=& m^2_1\cos^2\beta+m^2_2\sin^2\beta+\mathcal{G} T^2\equiv M^2_0+\mathcal{G} T^2, \\
c &=& R_1v_{S_1}+R_2v_{S_2},\quad \tilde{R}_\lambda=R_\lambda\sin\beta\cos\beta,\\
\tilde{\lambda}^2 &=& \frac{g^2_2+g^2_1}{8}\cos^22\beta+\frac{|\lambda|^2}{4}\sin^22\beta,
\end{eqnarray}
and we have subtracted $\boldsymbol{\rho}$-independent terms from $V(\boldsymbol{\rho}, T)$.  The value of $\mathcal{G}$ is given by the sum of the relevant couplings in the theory.  For simplicity, the temperature dependence in the mixing angle $\beta$ is neglected.  In order to have a stable vacuum in the symmetric phase, $m^2_S>0$ must hold.  Using the tadpole condition of $\rho_S$, $\rho_S$ can be written in terms of $\rho^2$ as
\begin{eqnarray}
\rho_S=\frac{c+\tilde{R}_\lambda\rho^2}{m^2_S+\frac{|\lambda|^2}{2}\rho^2},
\label{tad_rhoS}
\end{eqnarray}
which gives the trajectory of the minimum values of $\rho_S$ as a function of $\rho$.
Plugging this back into the Higgs potential (\ref{Veff_caseI}), we obtain
\begin{eqnarray}
  V(\rho, T)= \frac{1}{2}M^2(T)\rho^2
	-\frac{(c+\tilde{R}_\lambda\rho^2)^2}{2(m^2_S+\frac{|\lambda|^2}{2}\rho^2)}
	+\frac{\tilde{\lambda}^2}{4}\rho^4.\label{Vtree_sub}
\end{eqnarray}
The problem is now reduced to a one dimensional EWPT analysis.  The parameters $T_C$ and $\rho_C$ for the first order EWPT are found to be
\begin{eqnarray}
  T^2_C &=& \frac{F(\rho^2_C)-M^2_0}{a}=\frac{F(\rho^2_C)-F(v^2)}{a}, \\
  \rho^2_C &=& \frac{2}{|\lambda|^2}\left[-m^2_S+\frac{\sqrt{2m^2_S}}{\tilde{\lambda}}
	\bigg|\tilde{R}_\lambda-\frac{|\lambda|^2c}{2m^2_S}\bigg|\right].
\end{eqnarray}
Here, $F(\rho^2)$ is defined by
\begin{eqnarray}
F(\rho^2) = 2\tilde{R}_\lambda\left(\frac{c+\tilde{R}_\lambda\rho^2}
	{m^2_S+\frac{|\lambda|^2}{2}\rho^2}\right)
	-\frac{|\lambda|^2}{2}\left(\frac{c+\tilde{R}_\lambda\rho^2}
	{m^2_S+\frac{|\lambda|^2}{2}\rho^2}\right)^2-\tilde{\lambda}^2\rho^2.
\end{eqnarray} 
From $\rho^2_C>0$, we find the condition for the first order EWPT~\cite{Menon:2004wv} in terms of the model parameters:\footnote{$T^2_C>0$ also leads to the same condition.}
\begin{eqnarray}
  \tilde{\lambda}<\sqrt{\frac{2}{m^2_S}}
  \left|\tilde{R}_\lambda-\frac{|\lambda|^2c}{2m^2_S}\right|.
  \label{EWPT_nMSSM}
\end{eqnarray}
The physical implication of this condition becomes clearer if we expand the second term in Eq.~(\ref{Vtree_sub}) in powers of $\rho^2/m^2_S$.  To the sixth power of $\rho$, we obtain~\cite{Huber:2006wf}
\begin{eqnarray}
  V(\rho, T)&=&-\frac{c^2}{2m^2_S}
	+\frac{1}{2}\left\{M^2(T)-\frac{2c}{m^2_S}\left(\tilde{R}_\lambda
	-\frac{|\lambda|^2c}{4m^2_S}\right)\right\}\rho^2 \non\\
&&	+\frac{1}{4}\left\{\tilde{\lambda}^2-\frac{2}{m^2_S}\left(\tilde{R}_\lambda
	-\frac{|\lambda|^2c}{2m^2_S}\right)^2\right\}\rho^4
	+\frac{|\lambda|^2}{4m^4_S}\left(\tilde{R}_\lambda
	-\frac{|\lambda|^2c}{2m^2_S}\right)^2\rho^6.
	\label{expanded_V}
\end{eqnarray}
Condition (\ref{EWPT_nMSSM}) requires that the coefficient of the $\rho^4$-term be negative.  In other words, the roles of the negative cubic term and the positive quartic term in the usual scenario are now replaced by the negative quartic term and the positive sixth-power term.  As already pointed out in Ref.~\cite{Huber:2006wf}, the form of the Higgs potential (\ref{expanded_V}) is nothing but the SM Higgs potential with a dimension-six operator discussed in Refs.~\cite{SM_cutoff}.  
Hence the nMSSM can be regarded as one UV completion of such an effective theory.

It should be stressed that in order to make the first order EWPT stronger, $m^2_S$ should not be too large.  This implies some constraints on the Higgs bosons whose masses come from $m^2_S$.
Another important implication is the following.  From Eqs.~(\ref{tad_rhoS}) and (\ref{EWPT_nMSSM}), we find that to have a first order EWPT the difference between the singlet VEV in the broken phase and that in the symmetric phase
must be larger than some critical value
\begin{eqnarray}
  |\rho_S-\bar{\rho}_S|=\frac{\rho^2}{m^2_S}
	\frac{\left|\tilde{R}_\lambda-\frac{|\lambda|^2c}{2m^2_S}\right|}
	{1+\frac{|\lambda|^2\rho^2}{2m^2_S}}
	>\frac{1}{\sqrt{2m^2_S}}\frac{\tilde{\lambda}\rho^2}{1+\frac{|\lambda|^2\rho^2}{2m^2_S}},
\end{eqnarray}
where $\bar{\rho}_S=c/m^2_S$.  Conversely, if $\rho_S\simeq\bar{\rho}_S$, 
the singlet Higgs field will not play a significant role in realizing the first order EWPT, reducing to the MSSM-like EWPT.

In the sMSSM case, because of the presence of the $\rho^4_S$-term in the Higgs potential, the analytic formula for the first order EWPT is not as simple as the nMSSM case.  However, it is the same mechanism at work.  That is, the first order EWPT is possible due to the negative quartic term and the positive sixth-power term.

Although Eq.~(\ref{EWPT_nMSSM}) can provide a good approximation to the nMSSM~\cite{Huber:2006wf}, it fails to do so quantitatively in the sMSSM due to the presence of the secluded singlet Higgs bosons.  The precise value of $\bar{\rho}_S$ strongly depends on $\bar{\rho}_{S_i}$ through the tadpole conditions in the symmetric phase.  Indeed, it turns out that the above discussion is valid only qualitatively but not quantitatively.  We will present the numerical results in Sec.~\ref{sec:num_result}.
%
%
\subsection{Type B}\label{subsec:typeB}

As in the previous subsection, we focus only on the tree-level potential and the $\mathcal{O}(T^2)$ corrections coming from the finite-temperature effective potential.  As noted in subsection~\ref{subsec:pattern_EWPT},
$V$ can have an extremum at $(\rho_d, \rho_u, \rho_S, \rho_{S_1}, \rho_{S_2}) = (0,0,0,0,0)$ in Type B.  It is thus useful to prameterize the $\rho$-fields in terms of the five-dimensional polar coordinates
\begin{eqnarray}
\rho_d &=& z\cos\delta\cos\gamma\cos\alpha\cos\beta, \\
\rho_u &=& z\cos\delta\cos\gamma\cos\alpha\sin\beta, \\
\rho_S &=& z\cos\delta\cos\gamma\sin\alpha, \\
\rho_{S_1} &=& z\cos\delta\sin\gamma, \\
\rho_{S_2} &=& z\sin\delta.
\end{eqnarray}
Using these variables, the effective potential at $T_C$ takes the form
\begin{eqnarray}
V(z,T)=c_4z^4-c_3z^3+c_2z^2 = c_4z^2(z-z_C)^2,
\label{V_typeB}
\end{eqnarray}
where we have subtracted $z$-independent terms in Eq.~(\ref{V_typeB}), and
\begin{eqnarray}
z_C = \frac{c_3}{2c_4},\quad c_2 = \frac{c^2_3}{4c_4},
\end{eqnarray}
with
\begin{eqnarray}
  c_2 &=& \frac{1}{2}\Big[c^2_\delta c^2_\gamma c^2_\alpha
	(m^2_1c^2_\beta+m^2_2s^2_\beta)+m^2_Sc^2_\delta c^2_\gamma s^2_\alpha 
	+m^2_{S_1}c^2_\delta s^2_\gamma+m^2_{S_2}s^2_\delta \Big] \non\\
&&	-R_1c^2_\delta s_\gamma c_\gamma s_\alpha
	-R_2s_\delta c_\delta c_\gamma s_\alpha
	-(R_{12}+R_{\lambda_S}\rho_{S_3})s_\delta c_\delta s_\gamma
	 +\frac{|\lambda_S|^2}{4}(c^2_\delta s^2_\gamma+s^2_\delta)\rho^2_{S_3}\non\\	
&&	+\frac{g'^2_1}{4}Q_{S_3}\rho^2_{S_3}
	\Big[c^2_\delta c^2_\gamma c^2_\alpha
	(Q_{H_d}c^2_\beta+Q_{H_u}s^2_\beta)+Q_Sc^2_\delta s^2_\gamma s^2_\alpha
	+Q_{S_1}c^2_\delta s^2_\gamma+Q_{S_2}s^2_\delta\Big] \non\\
&&	+\frac{1}{2}\mathcal{G}c^2_\delta c^2_\gamma c^2_\alpha T^2_C,\\	
c_3 &=& R_\lambda c^3_\delta c^3_\gamma s_{\alpha} c^2_\alpha s_\beta c_\beta, \\
c_4 &=& \frac{g^2_2+g^2_1}{32}c^4_\delta c^4_\gamma c^4_\alpha c^2_{2\beta}
	+\frac{|\lambda|^2}{4}c^4_\delta c^4_\gamma c^2_\alpha
	(c^2_\alpha s^2_\beta c^2_\beta+s^2_\alpha) 
	+\frac{|\lambda_S|^2}{4}s^2_\delta c^2_\delta s^2_\gamma \non\\
&&	+\frac{g'^2_1}{8}\Big[c^2_\delta c^2_\gamma c^2_\alpha 
	+(Q_{H_d}c^2_\beta+Q_{H_u}s^2_\beta)+Q_Sc^2_\delta s^2_\gamma s^2_\alpha
	+Q_{S_1}c^2_\delta s^2_\gamma+Q_{S_2}s^2_\delta\Big]^2,
\end{eqnarray}
where $s_\alpha\equiv\sin\alpha, c_\alpha\equiv\cos\alpha$, etc
and the angles $(\alpha, \beta, \gamma, \delta)$ are evaluated at $T_C$.
Let us define $c_2 = k_1+k_2T_C^2$.  Then the critical temperature is given by
\begin{eqnarray}
T^2_C = \frac{1}{k_2}\left(\frac{c^2_3}{4c_4}-k_1\right).
\end{eqnarray}

In Type B, the magnitude of $R_\lambda$ is very important for the strong first order EWPT.  This is the same as the usual scenario in which the first order EWPT is induced by the negative cubic term.  Therefore, a large $m_{H^\pm}$ is favored.  We also note that a smaller $c_4$ is preferred for a larger $z_C$.  This implies that a light Higgs boson is favored.  We will quantify the statements here in Sec.~\ref{sec:num_result}.


\section{Numerical evaluations}\label{sec:num_result}

We give numerical results in this section.  To this end, $T_C$ and $\rho_C$ are determined using the one-loop effective potential at zero temperature, together with the fitting functions (\ref{V1_fit}) of $I_{B,F}(a^2)$.  First we discuss the $CP$-conserving case.  The case of $CP$ violation is argued in subsection \ref{subsec:CPV}.  Before showing the numerical results, we list the experimental constraints imposed in our numerical calculations.

For a Higgs boson lighter than $114.4$ GeV, we require
\begin{eqnarray}
\xi^2<k(m_{H_i}),
\label{LEP95excluded}
\end{eqnarray}
where $\xi=g_{H_iZZ}/g^{\rm SM}_{H_iZZ}$ and $k$ is the 95\% C.L. upper limit derived from the LEP experiments as a function of the Higgs boson mass.  For the $\rho$ parameter corrections, $\Delta\rho<2.0\times10^{-3}$ must be satisfied.  As mentioned briefly below Eq.~(\ref{Veff_1loop}), the constraints of the $Z'$ boson must be taken into account.  The $Z'$ boson mass ($m_{Z'}$) and the mixing between $Z$ and $Z'$ bosons ($\alpha_{ZZ'}$) are constrained by the direct searches of the $Z'$ boson and the EW precision measurements.  The typical values are found to be $m_{Z'}>1000$ GeV and $\alpha_{ZZ'}<\mathcal{O}(10^{-3})$~\cite{Zprime_search}.  Since we here consider the $\alpha_{ZZ'}=0$ case and at least one of the VEV's of the secluded singlets is taken to be $\mathcal{O}(1)$ TeV, the $Z'$ constraints are easily satisfied.  For the relevant SUSY particles, we impose $m_{\tilde{\chi}^\pm_1}>104$ GeV and $m_{\tilde{\chi}^0_1}>46$ GeV. 

In order to extract information about the doublet-singlet mixing effects, 
we define the MSSM fractions by~\cite{Han:2004yd}
\begin{eqnarray}
\xi_{H_i}=\left(O^{(H)}_{1i}\right)^2+\left(O^{(H)}_{2i}\right)^2,\quad \xi_{A_i}=\left(O^{(A)}_{1i}\right)^2,
\end{eqnarray}
where $O^{(H)}$ and $O^{(A)}$ are the orthogonal matrices which diagonalize the mass-squared matrices of the $CP$-even and $CP$-odd Higgs bosons, respectively.  The parameter $\xi$ characterizes to what extent $\phi~(=H,A)$ comprises the doublet components.  For $\xi=1$, $\phi$ are purely composed of the doublets while $\xi=0$ means that $\phi$ is purely the singlet components.
%
%
\subsection{$CP$-conserving case}

We turn off $CP$ violation in this and the next subsection.  As an example we take
\begin{eqnarray}
&&Q_{H_d}=Q_{H_u}=1,~\tan\beta=1,~|\lambda_S|=0.3,
~A_t=A_b=2m_{\tilde{q}}+\mu_{\rm eff}/\tan\beta, \non\\
&&{\rm sgn}(R_{\lambda_S})={\rm sgn}(R_{1,2})={\rm sgn}(R_{12})=+1,~
|m^2_{S_1S_2}|=(200~{\rm GeV})^2, \non\\
&&m_{\tilde{q}}=m_{\tilde{t}_R}=m_{\tilde{b}_R}=1000~{\rm GeV},~
M_1=200~{\rm GeV},~M_2=M'_1=300~{\rm GeV},
\end{eqnarray}
where $m_{\tilde{q}}$, $m_{\tilde{t}_R}$ and $m_{\tilde{t}_R}$ are soft SUSY breaking masses of the stop and sbottom, $M_1$, $M_2$ and $M'_1$ are the gaugino masses associated with $U(1)_Y$, $SU(2)_L$
and $U(1)'$, respectively.
The quantity ${\rm sgn}(R_\lambda)$ is fixed by
\begin{eqnarray}
{\rm sgn}(R_\lambda)={\rm sgn}\left(m^2_{H^\pm}-m^2_W+\frac{|\lambda|^2}{2}v^2
  -\Delta m^2_{H^\pm}\right),
\end{eqnarray}
where $\Delta m^2_{H^\pm}$ is the one-loop contribution to the charged Higgs boson mass.  

Since the mechanisms of the strong first order EWPT in Type A and Type B are different from 
each other qualitatively, we will explore both parameter spaces.
The following two cases are representative points for Types A and B, respectively.
\begin{eqnarray}
\mbox{Case 1}:&&v_S=500~{\rm GeV},~v_{S_1}=v_{S_2}=v_{S_3}=1200~{\rm GeV},~
	|A_{\lambda_S}|=|A_\lambda|, \\
\mbox{Case 2}:&&v_S=500~{\rm GeV},~v_{S_1}=v_{S_2}=100~{\rm GeV},~
v_{S_3}=1500~{\rm GeV},~|A_{\lambda_S}|=1000~{\rm GeV}.\non\\
\end{eqnarray}
As mentioned above, $|A_\lambda|$ is determined via the mass formula of $m_{H^\pm}$.
The remaining input parameters are $\lambda$, $m_{H^\pm}$, $|m^2_{SS_i}|, i=1,2$, by varying which we search for the strong first order EWPT.

\begin{figure}[t]
\begin{center}
\includegraphics[width=7cm]{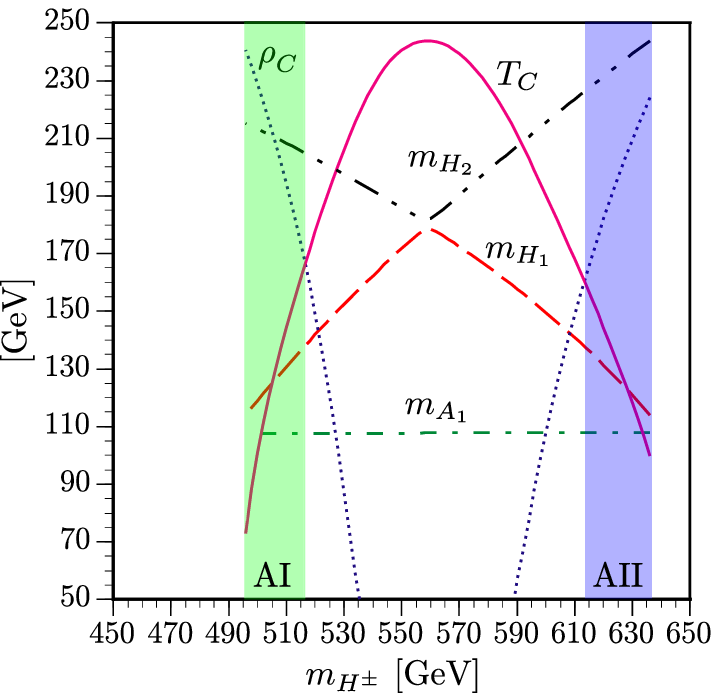}
\hspace{0.5cm}
\includegraphics[width=6.8cm]{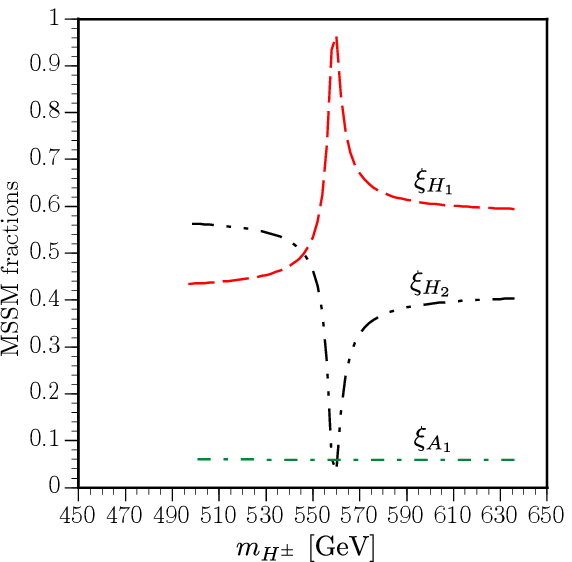}
\caption{Dependence of $\rho_C$, $T_C$, $m_{H_{1,2}}$, $m_{A_1}$ and the MSSM fractions of $H_1$, $H_2$ and $A_1$ on $m_{H^\pm}$ in Case 1 with $\lambda=0.8$ and $|m^2_{SS_1}|=|m^2_{SS_2}|=(50~{\rm GeV})^2$.}
\label{fig:case1_1}
\end{center}
\end{figure}

In the left panel of Fig.~\ref{fig:case1_1}, we plot $\rho_C$, $T_C$, $m_{H_{1,2}}$ and $m_{A_1}$ as a function of $m_{H^\pm}$ in Case 1.  As shown in Fig.~\ref{fig:Veff_tree}, the allowed region, where the prescribed EW vacuum is the global minimum, corresponds to the range, 496 GeV $\ltsim m_{H^\pm} \ltsim 636$ GeV.  It is found that there are two ranges in which $\rho_C/T_C>1$ is satisfied: $496$ GeV $\ltsim m_{H^\pm} \ltsim 516$ GeV (green-colored region) and $614$ GeV $\ltsim m_{H^\pm} \ltsim 636$ GeV (blue-colored region).  The strong first order EWPT is possible only for the region with sizable $|\Delta v_S|$; $|\Delta v_S|\gtsim 110$ GeV is obtained in Case 1.  In such a region, a low $T_C$ is enough for the EW symmetry restoration as expected.  We also plot the masses of $H_1$, $H_2$ and $A_1$.  To be consistent with the strong first order EWPT, the mass of the lightest $CP$-even Higgs boson $m_{H_1}$ should be less than about 130 GeV.

In the right panel of Fig.~\ref{fig:case1_1}, the MSSM fractions of $H_1$, $H_2$ and $A_1$ are plotted.  Deviations of $\xi_{H_{1,2}}$ from unity in the regions where the EWPT is strongly first order imply that the singlet Higgs bosons play a crucial role here.  Since $\xi_{A_1}<0.1$, $A_1$ is predominantly the singlet-like $CP$-odd Higgs boson.

\begin{figure}[t]
\begin{center}
\includegraphics[width=7cm]{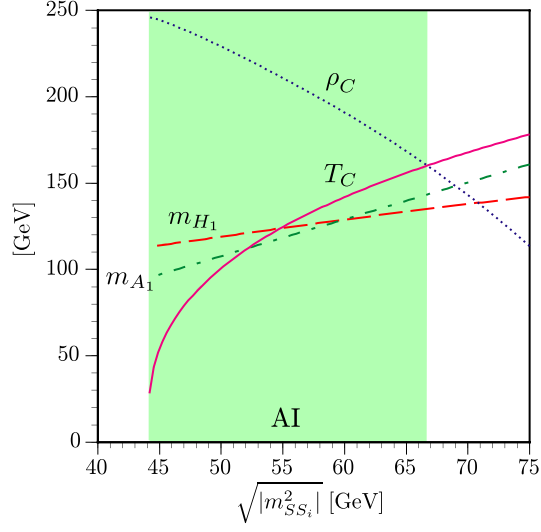}
\caption{Dependence of $\rho_C$, $T_C$, $m_{H_1}$ and $m_{A_1}$ on $\sqrt{|m^2_{SS_i}|}$, where $|m^2_{SS_i}| = |m^2_{SS_1}| = |m^2_{SS_2}|$ is assumed, in Case 1 with $\lambda=0.8$ and $m_{H^\pm}=500$ GeV.}
\label{fig:case1_2}
\end{center}
\end{figure}

Next we examine the dependences of $|m^2_{SS_1}|$ and $|m^2_{SS_2}|$ on $\rho_C/T_C$.  In Fig.~\ref{fig:case1_2}, we show $\rho_C$, $T_C$, $m_{H_1}$ and $m_{A_1}$ as a function of $\sqrt{|m^2_{SS_i}|}$, where $|m^2_{SS_i}| \equiv |m^2_{SS_1}| = |m^2_{SS_2}|$ is assumed.  As $|m^2_{SS_i}|$ increases, $\rho_C/T_C$ decreases.  Since a larger $|m^2_{SS_i}|$ gives a larger $m^2_S$ via the tadpole condition, this tendency is understandable from the discussions given in subsection.~\ref{subsec:typeA}.  Correspondingly, the Higgs boson masses which are mainly originated from $m^2_S$ are constrained, leading to the mass bounds: $m_{H_1}\ltsim 135$ GeV and $m_{A_1}\ltsim 143$ GeV to have the strong first order EWPT.

\begin{figure}[t]
\begin{center}
\includegraphics[width=7cm]{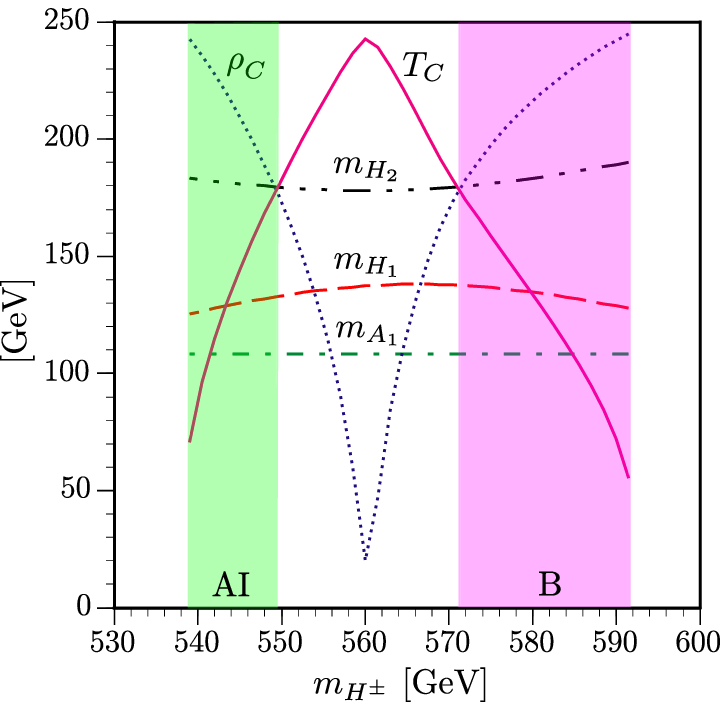}
\hspace{0.5cm}
\includegraphics[width=6.8cm]{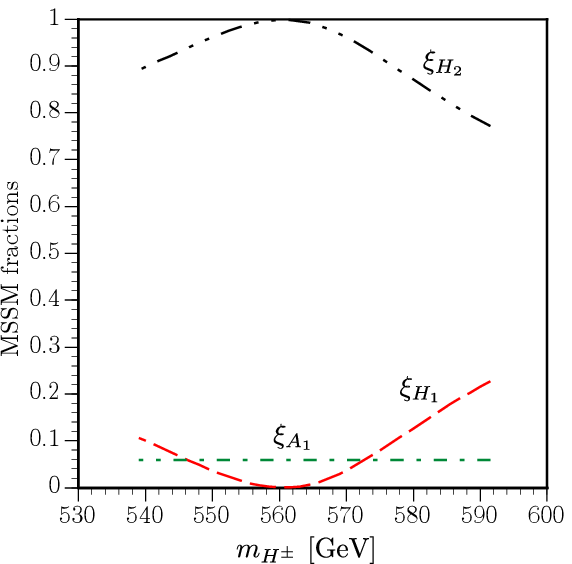}
\caption{Dependence of $\rho_C$, $T_C$, $m_{H_{1,2}}$ and $m_{A_1}$ (left panel) and 
$\xi_{H_{1,2}}$ and $\xi_{A_1}$ (right panel) on $m_{H^\pm}$ in Case 2 with $\lambda=0.8$, $|m^2_{SS_1}|=|m^2_{SS_2}|=(200~{\rm GeV})^2$.}
\label{fig:case2_1}
\end{center}
\end{figure}

The numerical results for Case 2 are shown in Fig.~\ref{fig:case2_1}.  As done with Case 1, we plot $\rho_C$, $T_C$, $m_{H_{1,2}}$ and $m_{A_1}$ (left panel) and $\xi_{H_{1,2}}$ and $\xi_{A_1}$ (right panel) as a function of $m_{H^\pm}$.  Similar to Case 1, due to the vacuum condition the allowed region is limited to the relatively small range $539$ GeV $\ltsim m_{H^\pm} \ltsim 592$ GeV.  Within the interval, there are two regions where the strong first order EWPT is possible: $539$ GeV $\ltsim m_{H^\pm} \ltsim 548$ GeV (green-colored region) and $572$ GeV $\ltsim m_{H^\pm} \ltsim 592$ GeV (magenta-colored region).  The former corresponds to Type AI EWPT and the latter to Type B EWPT.  For Type B, as discussed in subsection~\ref{subsec:typeB}, the magnitude of $R_\lambda$,
and thus that of $m_{H^\pm}$ is crucial for the strength of the first order EWPT.  As $m_{H^\pm}$ increases, $\rho_C/T_C$ becomes larger.  From the right panel of Fig.~\ref{fig:case2_1}, we find that the doublet-singlet mixing plays an essential role in realizing the strong first order EWPT as it should be.  The mass limits of $H_1$ and $A_1$ consistent with $\rho_C/T_C>1$ are found to be, respectively, $m_{H_1}\ltsim 137$ GeV and $m_{A_1}\ltsim 108$ GeV for this specific parameter set.


\subsection{Scan analysis}

In order to search for strong first order EWPT in the wider parameter space, we perform scans in the $\lambda$-$m_{H^\pm}$ and $\sqrt{|m^2_{SS_i}|}$-$m_{H^\pm}$ planes, respectively.

\begin{figure}[t]
\begin{center}
\includegraphics[width=6.8cm]{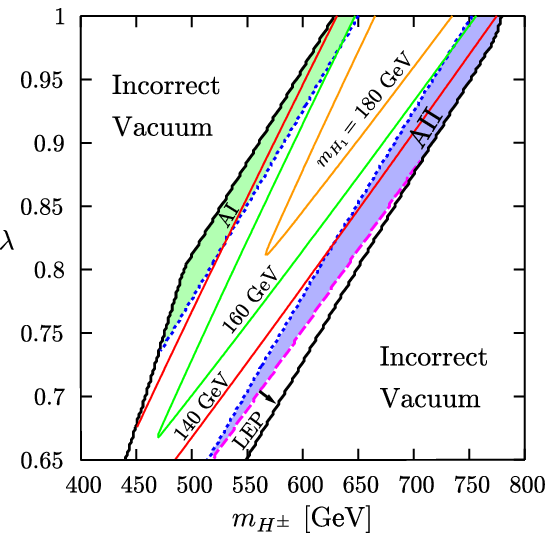}
\hspace{0.5cm}
\includegraphics[width=7cm]{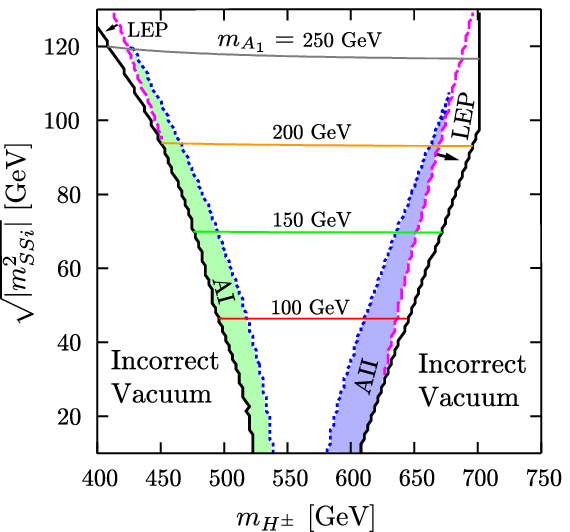}
\caption{Left: Case 1 with $|m^2_{SS_1}|=|m^2_{SS_2}|=(50~{\rm GeV})^2$ in the 
$\lambda$-$m_{H^\pm}$ plane.
Right: Case 1 with $\lambda=0.8$ and $|m^2_{SS_1}|=|m^2_{SS_2}|$ in the $\sqrt{|m^2_{SS_i}|}$-$m_{H^\pm}$ plane.}
 \label{fig:case1_3}
\end{center}
\end{figure}

We show the results for Case 1 in Fig.~\ref{fig:case1_3}.  In the left plot, we take $|m^2_{SS_1}|=|m^2_{SS_2}|=(50~{\rm GeV})^2$ and show the regions with sufficiently strong first order EWPT in the $\lambda$-$m_{H^\pm}$ plane.  In the area surrounded by the solid black curves, the prescribed EW vacuum is the global minimum.  The region below the magenta dashed line is excluded by the LEP 95\% C.L. exclusion limit, Eq.~(\ref{LEP95excluded}).  The green (blue) region corresponds to Type AI (AII) EWPT.  We also overlay the contours of the mass of the lightest Higgs boson, $m_{H_1}=140$ GeV (red), $160$ GeV (green) and $180$ GeV (orange).  Around the $\lambda=1$ region, $m_{H_1}=140$ to $160$ GeV is consistent with the strong first order EWPT.  Here, we do not impose the perturbativity of $\lambda$ in the investigation.  However, the maximally allowed $m_{H_1}$ satisfying $\rho_C/T_C>1$ is expected to be larger than that of the MSSM.  We will compare our results with the other models in subsection~\ref{subsec:comparison}.

In the right plot, we take Case 1 with $\lambda=0.8$ in the $\sqrt{|m^2_{SS_i}|}$-$m_{H^\pm}$ plane and $|m^2_{SS_1}|=|m^2_{SS_2}|$ is assumed.  The curves have the same meanings as the left plot except that the Higgs boson mass contours are drawn for $A_1$.  Here, we plot for $m_{A_1}=100$ GeV (red), $150$ GeV (green), $200$ GeV (orange) and $250$ GeV (gray).  It should be emphasized that $m_{A_1}<250$ GeV in order to be consistent with the strong first order EWPT.  This feature differs from the MSSM.

\begin{figure}[t]
\begin{center}
\includegraphics[width=6.8cm]{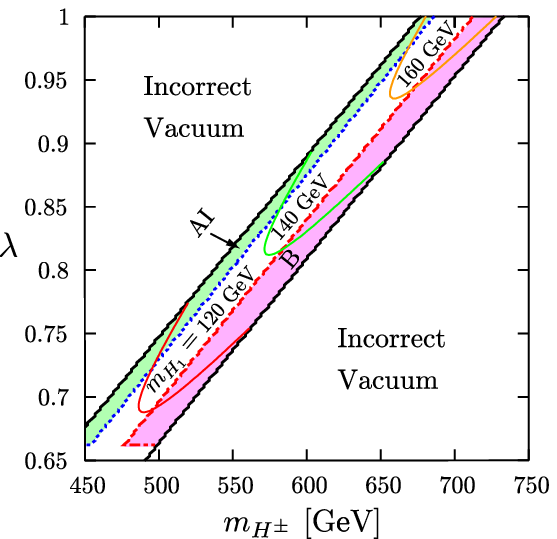}
\hspace{0.5cm}
\includegraphics[width=7cm]{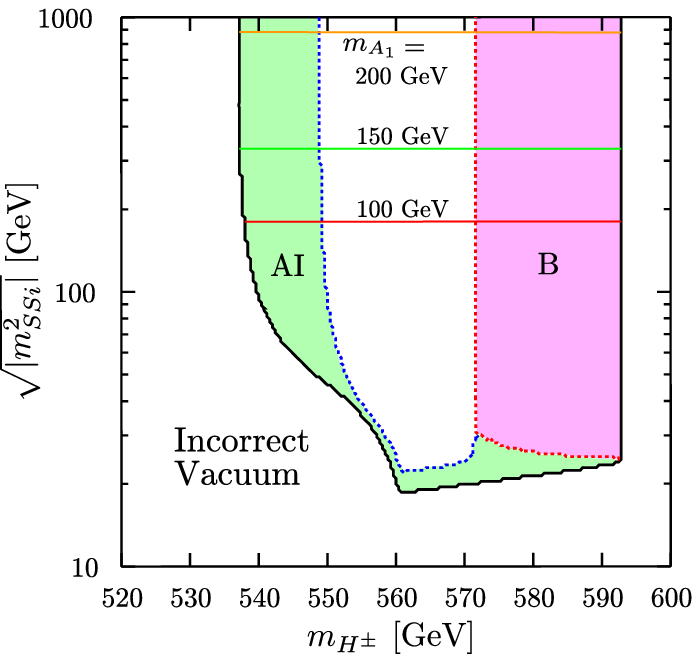}
\caption{Left: Case 2 with $|m^2_{SS_1}|=|m^2_{SS_2}|=(200~{\rm GeV})^2$ in the 
$\lambda$-$m_{H^\pm}$ plane.
Right: Case 2 with $\lambda=0.8$ and $|m^2_{SS_1}|=|m^2_{SS_2}|$ in the $\sqrt{|m^2_{SS_i}|}$-$m_{H^\pm}$ plane.}
\label{fig:case2_2}
\end{center}
\end{figure}

The numerical results of Case 2 is shown in Fig.~\ref{fig:case2_2}.  In the left panel, we take $|m^2_{SS_1}|=|m^2_{SS_2}|=(200~{\rm GeV})^2$ and show the regions with sufficiently strong first order EWPT in the $\lambda$-$m_{H^\pm}$ plane.  The magenta region represents Type B EWPT.  The meanings of the remaining curves are the same as in Case 1.  The $\lambda$ dependence of Type B EWPT is expected from the analysis done in subsection~\ref{subsec:typeB}.  Since $c_3\propto \lambda$ and $c_4\propto \lambda^2$ in the potential (\ref{V_typeB}), $\rho_C/T_C$ becomes smaller as $\lambda$ increases.  In both Type A and Type B EWPT, it is observed that $m_{H_1}=160$ GeV supports the strong first order EWPT if $\lambda \simeq 1$.

In the right panel, we take $\lambda=0.8$ in the $\sqrt{|m^2_{SS_i}|}$-$m_{H^\pm}$ plane, and $|m^2_{SS_1}|=|m^2_{SS_2}|\equiv|m^2_{SS_i}|$ is assumed.  In Case 2, the dependence of $|m^2_{SS_i}|$ on the strength of the first order EWPT is smaller. This is because both $v_{S_1}$ and $v_{S_2}$ are taken to be small (100 GeV) so that the variation of $m^2_S$ as a function of $|m^2_{SS_i}|$ is much milder than that in Case 1.  The $m_{A_1}$ contours are drawn in the red curve (100 GeV), the green curve (150 GeV), and the orange curve (200 GeV), respectively.  Asymptotically, $m_{A_1}$ approaches around 214 GeV as $|m^2_{SS_i}|$ increases.  Therefore, there is no significant difference in the Higgs mass spectrum between Case 1 and Case 2.

Here, we comment on the other cases.  Although the magnitude of $v_S$ is relevant to the realization of the non-MSSM-like EWPT, so that it should be taken around $\mathcal{O}(100)$ GeV, the maximal value of $\rho_C/T_C$ is not sensitive to its exact value.  Similarly, the strength of the first order EWPT does not depend sensitively on the other unvaried parameters appearing in the tree-level Higgs potential.  Therefore, the upper limits of the Higgs boson masses allowed for the strong first order EWPT would not change substantially.

Even in the case of the non-MSSM-like EWPT, we could also argue that the light stop is lighter than the top quark. In Case 1 with the light stop, no significant enhancement on the strength of the first order EWPT is observed.  It implies that the singlet Higgs bosons and the light stop do not contribute to the first order EWPT constructively.  This may follow from the fact that the light stop gives a $-\rho^3$-like term in the effective potential while the singlet Higgs bosons produces a $-\rho^4$-like term.  In Case 2 with the light stop, however, the stop contribution is constructive, leading to about 60\% enhancement on $\rho_C/T_C$.  Since the right-handed soft SUSY breaking mass and the off-diagonal term in the stop mass matrix are taken to be small in the light stop scenario, the upper bound of $m_{H_1}$ obtained above is virtually unchanged.


\subsection{$CP$-violating case}\label{subsec:CPV}

We now discuss $CP$ violation in this subsection.  The relations between the $CP$-violating phases at the one-loop level are given by the one-loop tadpole conditions~\cite{Chiang:2008ud}:
\begin{eqnarray}
  I_\lambda&=&-\frac{N_C}{8\pi^2 v^2}
	\left[\frac{m^2_tI_t}{\sin^2\beta}f(m^2_{\tilde{t}_1},m^2_{\tilde{t}_2})
	+\frac{m^2_bI_b}{\cos^2\beta}f(m^2_{\tilde{b}_1},m^2_{\tilde{b}_2})\right], \\
I_{\lambda_S}&=&0,\quad
I_1=I_{12}\frac{v_{S_2}}{v_S}, \quad
I_2=-I_{12}\frac{v_{S_1}}{v_S},
\label{tad_CPV}
\end{eqnarray}
where $I_q={\rm Im}(\lambda A_q)/\sqrt{2},~q=t,b$.  Here $A_q$ is the soft SUSY-breaking trilinear coupling, and $f(m^2_1,m^2_2)$ is defined by
\begin{eqnarray}
f(m^2_1,m^2_2) = \frac{1}{m^2_1-m^2_2}
	\left[m^2_1\left(\ln\frac{m^2_1}{M^2}-1\right)
	-m^2_2\left(\ln\frac{m^2_2}{M^2}-1\right)\right].
\end{eqnarray}
To see the $CP$-violating effect that the MSSM does not possess, we take $I_t=I_b=0$ so that $I_\lambda=0$.  It follows from Eq.~(\ref{tad_CPV}) that
\begin{eqnarray}
  \sin\theta_{SS_1}&=&\left|\frac{m^2_{S_1S_2}}{m^2_{SS_1}}\right|\frac{v_{S_2}}{v_S}
	\sin\theta_{S_1S_2}, \label{tad_CPV1}\\
\sin\theta_{SS_2}&=&-\left|\frac{m^2_{S_1S_2}}{m^2_{SS_2}}\right|\frac{v_{S_1}}{v_S}
	\sin\theta_{S_1S_2}\label{tad_CPV2},
\end{eqnarray}
where $\theta_{SS_{1,2}}={\rm Arg}(m^2_{SS_{1,2}})$.  In the $CP$-violating case, the prefactors in Eqs.~(\ref{tad_CPV1}) and (\ref{tad_CPV2}) cannot be chosen arbitrarily.  For instance, the following inequalities must hold
for $\sin\theta_{S_1S_2}=1$:
\begin{eqnarray}
|m^2_{S_1S_2}|\frac{v_{S_2}}{v_S}\leq |m^2_{SS_1}|,\quad
|m^2_{S_1S_2}|\frac{v_{S_1}}{v_S}\leq|m^2_{SS_2}|.
\label{ineq_mSSi}
\end{eqnarray}
As we have discussed so far, small $|m^2_{SS_{1,2}}|$ are favored for the strong first order EWPT in Case 1, which in turn imposes constraints on the magnitude of $|m^2_{S_1S_2}|$ via Eq.~(\ref{ineq_mSSi}).  Since $|m^2_{S_1S_2}|$ appears in some of the diagonal elements of the neutral Higgs bosons, {\it i.e.}, the $(h_{S_{1,2}},h_{S_{1,2}})$ and $(a_{S_{1,2}},a_{S_{1,2}})$ elements~\cite{Chiang:2008ud}, the Higgs boson masses mainly coming from those elements are lowered in accordance with $|m^2_{S_1S_2}|$.

\begin{figure}
\begin{center}
\includegraphics[width=6.5cm]{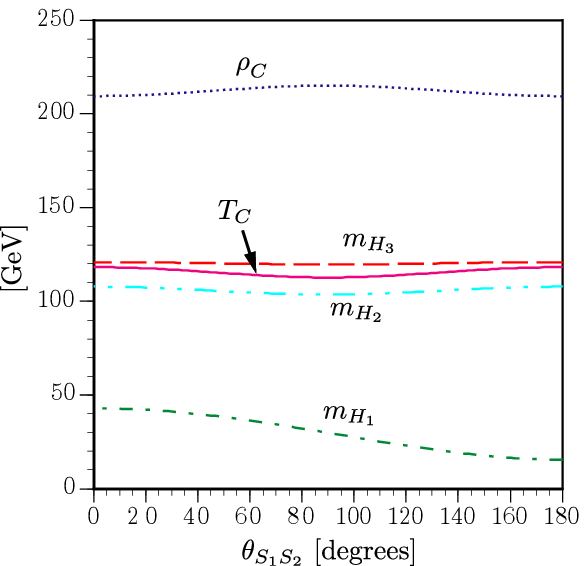}
\hspace{1cm}
\includegraphics[width=6.5cm]{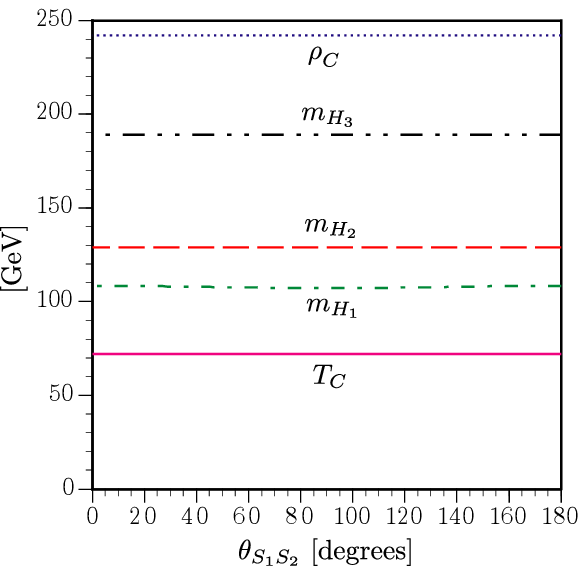}
\caption{Left: Case 1 with $|\lambda|=0.8$, $m_{H^\pm}=630$ GeV, 
$|m^2_{SS_1}|=|m^2_{SS_2}|=(50~{\rm GeV})^2$,
and $|m^2_{S_1S_2}|=(20~{\rm GeV})^2$. 
Right: Case 2 with $|\lambda|=0.8$, $m_{H^\pm}=590$ GeV, and
$|m^2_{SS_1}|=|m^2_{SS_2}|=(200~{\rm GeV})^2$.}
\label{fig:CPV}
\end{center}
\end{figure}

In Fig.~\ref{fig:CPV}, we plot $\rho_C$, $T_C$, and the three Higgs boson masses $m_{H_i}, i=1,2,3,$ for Case 1 with $|\lambda|=0.8$, $m_{H^\pm}=630$ GeV, $|m^2_{SS_1}|=|m^2_{SS_2}|=(50~{\rm GeV})^2$ and $|m^2_{S_1S_2}|=(20~{\rm GeV})^2$ (left panel); and for Case 2 with $|\lambda|=0.8$, $m_{H^\pm}=590$ GeV, and $|m^2_{SS_1}|=|m^2_{SS_2}|=(200~{\rm GeV})^2$ (right panel).  In both cases, the $CP$-violating effect on the strength of the first EWPT is relatively mild.  In particular, no significant dependence is observed in Case 2.  This is because the mass matrix elements of the $CP$-even and -odd mixing part are not sufficiently large to alter the Higgs boson masses significantly.

Since $|m^2_{S_1S_2}|$ assumes a smaller value to be consistent with Eq.~(\ref{ineq_mSSi}) in Case 1, some of the Higgs masses are lowered as seen in the left panel of Fig.~\ref{fig:CPV}.  In this case, $H_1$ and $H_2$ are purely singlet-like Higgs bosons and may be challenging to detect at colliders.

In contrast to Case 1, such a small $|m^2_{S_1S_2}|$ is not mandatory in Case 2 since $v_{S_1}$ and $v_{S_1}$ are small enough to satisfy Eq.~(\ref{ineq_mSSi}).  Therefore, there is no significant difference between the Higgs mass spectrum of the $CP$-conserving case and that of the $CP$-violating case.

\begin{figure}
\begin{center}
\includegraphics[width=6.5cm]{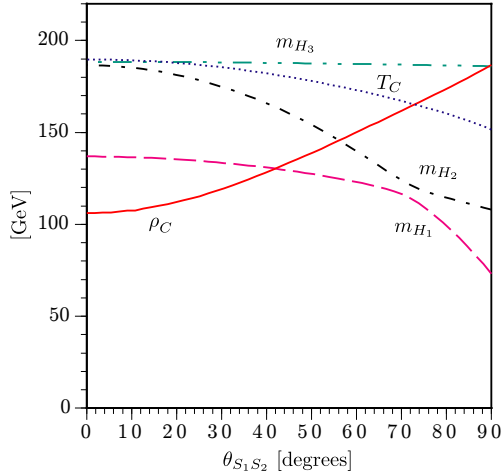}
\caption{Dependence of various quantities on the CP-violating phase $\theta_{S_1S_2}$.  Here we take $\lambda=0.8, m_{H^\pm}=675$ GeV, $v_S=560$ GeV, 
$v_{S_1}=v_{S_3}=100$ GeV, $v_{S_2}=1000$ GeV, $|m^2_{SS_1}|=(334~{\rm GeV}^2)$,
$|m^2_{SS_2}|=(106~{\rm GeV}^2)$, and $|m^2_{S_1S_2}|=(200~{\rm GeV}^2)$.}
\label{fig:case3}
\end{center}
\end{figure}

Depending on the choices of the Higgs VEVs, one can possibly find cases that have a significant dependence on $\theta_{S_1S_2}$.  One example is shown in Fig.~\ref{fig:case3}.  Here we take $|\lambda|=0.8, m_{H^\pm}=675$ GeV, $v_S=560$ GeV, $v_{S_1}=v_{S_3}=100$ GeV, $v_{S_2}=1000$ GeV, $|m^2_{SS_1}|=(334~{\rm GeV})^2$, $|m^2_{SS_2}|=(106~{\rm GeV})^2$, and $|m^2_{S_1S_2}|=(200~{\rm GeV})^2$.  In this case, a stable EW vacuum exists only for $\theta_{S_1S_2}\ltsim 90^\circ$, beyond which it shifts to a metastable one.  It is found that $\rho_C/T_C<1$ at $\theta_{S_1S_2}=0$.  As $\theta_{S_1S_2}$ increases, however, $\rho_C/T_C$ gets enhanced and eventually $\rho_C/T_C>1$ is fulfilled for $\theta_{S_1S_2}\gtsim 73.8^\circ$. This example clearly illustrates that CP violation in the model can enhance the first order EWPT. It should be noted that this behavior may not be observed in the MSSM, for the strong first order EWPT favors vanishing off-diagonal elements of the stop mass matrix, where the $CP$-violating phase resides.  On the contrary, such a $CP$-violating phase can make the EWPT weaker in the MSSM~\cite{Funakubo:2002yb}.

Finally, we briefly comment on the constraints from the electric dipole moment (EDM).  Here, we only consider CP violation originating from the soft SUSY breaking masses $m^2_{SS_1}$, $m^2_{SS_2}$ and $m^2_{S_1S_2}$.  In such a case, the contributions to the EDM's of electron and neutron and so on are small enough to satisfy the current experimental limits~\cite{Kang:2009rd}.


\subsection{Comparisons}\label{subsec:comparison}

In this subsection, we compare our results with other models.  The mass spectra having the strong first order EWPT are summarized in Table~\ref{comparison}.  In the sMSSM, $m_{H_1}\ltsim 160$ GeV and $m_{A_1}\ltsim 250$ GeV are required.  However, there is no constraint on the stop mass.  It is well-known that in the MSSM, in addition to the requirement of a light Higgs boson, the mass of the lighter stop must be smaller than that of top quark.  According to Ref.~\cite{Carena:2008vj}, $m_{H_1}\ltsim 127$ GeV is required.  An even severer bound $m_{\tilde{t}_1}<120$ GeV is found to be required by the strong first order EWPT.

In the NMSSM, it is claimed that $m_{H_1}\simeq170$ GeV is compatible with the strong first order EWPT~\cite{Pietroni:1992in}.  In Ref~\cite{Pietroni:1992in}, $m_{\tilde{q}}=m_{\tilde{t}}=1000$ GeV are taken, the two stop masses are assumed to be degenerate, and the perturbativity of $\lambda$ is taken into account.  The sphaleron decoupling condition is $\rho_C/T_C\gtsim 1.3$.  Here, $T_C$ is defined as the temperature at which the effective potential at the origin is destabilized in some direction.  Such a temperature is always lower than the temperature that we define in this paper, rendering a larger $\rho_C/T_C$.

In the nMSSM, the $CP$-odd Higgs boson must have an upper bound on its mass in order to realize the strong first order EWPT as discussed in subsection \ref{subsec:typeA}.  In Ref.~\cite{Menon:2004wv}, such an upper limit is found to be $m_{A_1}\ltsim 250$ GeV, which is approximately the same as the sMSSM.

In the columns of NMSSM, nMSSM and UMSSM in Table~\ref{comparison}, we leave a question mark for those cases where we are not aware of any literature that gives an upper bound on the Higgs boson mass consistent with the requirement of strong first order EWPT.  As a reference, we cite the value of the $CP$-even Higgs boson in the nMSSM from Ref.~\cite{Menon:2004wv}.  It is found that the lightest $CP$-even Higgs boson with a mass of 130 GeV is consistent with the strong first order EWPT.  The main differences between our work and Ref.~\cite{Menon:2004wv} are the following: (1) The soft SUSY-breaking parameters are taken as $m_{\tilde{q}}=m_{\tilde{t}}=500$ GeV and $A_t=100$ GeV.  (2) The perturbativity of $\lambda~(\ltsim 0.8)$ is imposed.  (3) The sphaleron decoupling condition is $\rho_C/T_C\gtsim1.3$.  When we na{\"i}vely adopt those three conditions in the sMSSM, the upper bound on $m_{H_1}$ in the sMSSM approaches around 130 GeV.

In the NMSSM~\cite{Huber:2000mg} and the UMSSM~\cite{EWPT_UMSSM}, it is found that the strength of the first order EWPT can be enhanced for a larger mass of the lightest Higgs boson.  In contrast, we do not observe such an effect in the sMSSM.

\begin{table}[t]
\begin{center}
\begin{tabular}{|c|c|c|c|c|c|}
\hline
 & sMSSM & MSSM & NMSSM & nMSSM & UMSSM \\
 \hline
$m_{H_1}$ & $\ltsim160$ GeV & $\ltsim 127$ GeV~\cite{Carena:2008vj}
	& $\ltsim170$ GeV~\cite{Pietroni:1992in} & ? & ? \\
$m_{A_1}$ & $\ltsim250$ GeV & --- & ? & $\ltsim250$ GeV~\cite{Menon:2004wv} & ? \\
$m_{\tilde{t}_1}$ & --- & $\ltsim120$ GeV~\cite{Carena:2008vj} $<m_t$ & --- & --- & --- \\
\hline
\end{tabular}
\end{center}
\caption{The mass spectra for the strong first order EWPT in the various models.}
\label{comparison}
\end{table}


\section{Conclusions and Discussions}\label{sec:conclusion}

We have investigated the possible parameter space where first order EWPT is possible and strong enough for successful EW baryogenesis in the sMSSM.  We demonstrate two typical examples.  In Case 1 where all of the secluded singlet Higgs VEV's are taken to be of the order of TeV, Type A EWPT is realized.  This pattern of the EWPT is the same as in the nMSSM in the sense that the EWPT is possible by the non-cubic coupling with a negative coefficient.  However, the dependences of the strength of the EWPT on $m_{H^\pm}$ or $A_\lambda$ in both models are different from each other because of the $U(1)'$ contributions and the secluded singlet sector.  In Case 2 where two of the secluded singlet Higgs VEV's are taken to be $\mathcal{O}(100)$ GeV, Type B EWPT is realized.  In Type B, the parameters most relevant to the strong first order EWPT are $m_{H^\pm}$ and $\lambda$.  As $m_{H^\pm}$ increases, $\rho_C/T_C$ is enhanced.  However, as $\lambda$ increases, the magnitude of $\rho_C/T_C$ reduces.  The mechanism of strong first order EWPT in Type B is quite similar to the usual case where one has negative cubic and positive quartic terms.

By scanning the parameters most relevant to the EWPT, we have obtained the typical Higgs mass spectrum that is consistent with the strong first order EWPT.  In the sMSSM, it is found that $m_{H_1}\ltsim 160$ GeV and $m_{A_1}\ltsim 250$ GeV must be satisfied.  Unlike the MSSM, the light stop mass is not necessarily smaller than the top quark mass.

We have also worked out the impact of the $CP$-violating phase on the strength of the first order EWPT.  In most of the parameter space, the dependence of $\rho_C/T_C$ on $\theta_{S_1S_2}$ is quite mild.  However, according to the tadpole conditions for the $CP$-odd Higgs bosons, $|m^2_{SS_{1}}|$, $|m^2_{SS_{2}}|$ and $|m^2_{S_1S_2}|$, $v_S$ and $v_{S_{1,2}}$ cannot be freely chosen.  This leads to the constraints on the Higgs mass spectrum, especially for Case 1.

Our numerical study suggests that to have non-MSSM-like EWPT, the singlet Higgs VEV's, particularly $v_S$, in the broken phase and the symmetric phase must be significantly different from each other.  For the typical parameter sets, $|\Delta v_S|>100$ GeV must be satisfied.  This condition is almost temperature-independent in our analysis.  In principle, $\Delta v_S$ can be derived provided that the soft SUSY-breaking masses are known, or more precisely, once the global structure of the Higgs potential is completely determined.  The determination of a sizable $|\Delta v_S|$ at zero temperature from collider experiments may be evidence of strong first order EWPT in the singlet-extended MSSM.

\appendix

\begin{acknowledgments}
E.~S. would like to thank Koichi Funakubo for useful discussions.  This work was partly carried out while E.~S. visited KEK under the KEK-NCTS Exchange Program.  This work is supported in part by the National Science Council of Taiwan, R.~O.~C.\ under Grant No.~NSC~97-2112-M-008-002-MY3 and in part by the NCTS.
\end{acknowledgments}



\begin{thebibliography}{99}

\bibitem{Amsler:2008zzb}
  C.~Amsler {\it et al.}  [Particle Data Group],
  ``Review of particle physics,''
  Phys.\ Lett.\  B {\bf 667} (2008) 1.

\bibitem{Sakharov:1967dj}
  A.~D.~Sakharov,
  ``Violation of CP Invariance, c Asymmetry, and Baryon Asymmetry of the
  Universe,''
  Pisma Zh.\ Eksp.\ Teor.\ Fiz.\  {\bf 5} (1967) 32
  [JETP Lett.\  {\bf 5} (1967\ SOPUA,34,392-393.1991\ UFNAA,161,61-64.1991) 24].

\bibitem{CKM}
  N.~Cabibbo,
  ``Unitary Symmetry and Leptonic Decays,''
  Phys.\ Rev.\ Lett.\  {\bf 10} (1963) 531;

  M.~Kobayashi and T.~Maskawa,
  ``CP Violation In The Renormalizable Theory Of Weak Interaction,''
  Prog.\ Theor.\ Phys.\  {\bf 49} (1973) 652.

\bibitem{ewbg_sm_cp}
  M.~B.~Gavela, P.~Hernandez, J.~Orloff and O.~Pene,
  ``Standard Model CP-violation and Baryon asymmetry,''
  Mod.\ Phys.\ Lett.\  A {\bf 9} (1994) 795
  [arXiv:hep-ph/9312215];

  M.~B.~Gavela, P.~Hernandez, J.~Orloff, O.~Pene and C.~Quimbay,
  ``Standard model CP violation and baryon asymmetry. Part 2: Finite
  temperature,''
  Nucl.\ Phys.\  B {\bf 430} (1994) 382
  [arXiv:hep-ph/9406289];

  P.~Huet and E.~Sather,
  ``Electroweak baryogenesis and standard model CP violation,''
  Phys.\ Rev.\  D {\bf 51} (1995) 379
  [arXiv:hep-ph/9404302];

  T.~Konstandin, T.~Prokopec and M.~G.~Schmidt,
  ``Axial currents from CKM matrix CP violation and electroweak
  baryogenesis,''
  Nucl.\ Phys.\  B {\bf 679} (2004) 246
  [arXiv:hep-ph/0309291].

\bibitem{crossover}
  K.~Kajantie, M.~Laine, K.~Rummukainen and M.~E.~Shaposhnikov,
  ``Is there a hot electroweak phase transition at m(H) > approx. m(W)?,''
  Phys.\ Rev.\ Lett.\  {\bf 77} (1996) 2887
  [arXiv:hep-ph/9605288];

  K.~Rummukainen, M.~Tsypin, K.~Kajantie, M.~Laine and M.~E.~Shaposhnikov,
  ``The universality class of the electroweak theory,''
  Nucl.\ Phys.\  B {\bf 532} (1998) 283
  [arXiv:hep-lat/9805013];

  F.~Csikor, Z.~Fodor and J.~Heitger,
  ``Endpoint of the hot electroweak phase transition,''
  Phys.\ Rev.\ Lett.\  {\bf 82} (1999) 21
  [arXiv:hep-ph/9809291];

  Y.~Aoki, F.~Csikor, Z.~Fodor and A.~Ukawa,
  ``The endpoint of the first-order phase transition of the SU(2)  gauge-Higgs
  model on a 4-dimensional isotropic lattice,''
  Phys.\ Rev.\  D {\bf 60} (1999) 013001
  [arXiv:hep-lat/9901021].

\bibitem{ewbg}
For reviews, see
A.~G.~Cohen, D.~B.~Kaplan and A.~E.~Nelson,
``Progress in electroweak baryogenesis,''
Ann.\ Rev.\ Nucl.\ Part.\ Sci.\  {\bf 43} (1993) 27
[arXiv:hep-ph/9302210];

M.~Quiros,
``Field theory at finite temperature and phase transitions,''
Helv.\ Phys.\ Acta {\bf 67} (1994) 451;

V.~A.~Rubakov and M.~E.~Shaposhnikov,
``Electroweak baryon number non-conservation in the early universe and in
high-energy collisions,''
Usp.\ Fiz.\ Nauk {\bf 166} (1996) 493
[Phys.\ Usp.\  {\bf 39} (1996) 461]
[arXiv:hep-ph/9603208];

K.~Funakubo,
``CP violation and baryogenesis at the electroweak phase transition,''
Prog.\ Theor.\ Phys.\  {\bf 96} (1996) 475
[arXiv:hep-ph/9608358];

M.~Trodden,
``Electroweak baryogenesis,''
Rev.\ Mod.\ Phys.\  {\bf 71} (1999) 1463
[arXiv:hep-ph/9803479];

W.~Bernreuther,
``CP violation and baryogenesis,''
Lect.\ Notes Phys.\  {\bf 591} (2002) 237
[arXiv:hep-ph/0205279].

\bibitem{ewbg-mssm}
  M.~S.~Carena, M.~Quiros and C.~E.~M.~Wagner,
  ``Opening the Window for Electroweak Baryogenesis,''
  Phys.\ Lett.\  B {\bf 380} (1996) 81
  [arXiv:hep-ph/9603420];

  D.~Delepine, J.~M.~Gerard, R.~Gonzalez Felipe and J.~Weyers,
  ``A light stop and electroweak baryogenesis,''
  Phys.\ Lett.\  B {\bf 386} (1996) 183
  [arXiv:hep-ph/9604440];

  P.~Huet and A.~E.~Nelson,
  ``Electroweak baryogenesis in supersymmetric models,''
  Phys.\ Rev.\  D {\bf 53} (1996) 4578
  [arXiv:hep-ph/9506477];

  B.~de Carlos and J.~R.~Espinosa,
  ``The baryogenesis window in the MSSM,''
  Nucl.\ Phys.\  B {\bf 503} (1997) 24
  [arXiv:hep-ph/9703212];

  M.~S.~Carena, M.~Quiros, A.~Riotto, I.~Vilja and C.~E.~M.~Wagner,
  ``Electroweak baryogenesis and low energy supersymmetry,''
  Nucl.\ Phys.\  B {\bf 503} (1997) 387
  [arXiv:hep-ph/9702409];

  M.~S.~Carena, M.~Quiros and C.~E.~M.~Wagner,
  ``Electroweak baryogenesis and Higgs and stop searches at LEP and the
  Tevatron,''
  Nucl.\ Phys.\  B {\bf 524} (1998) 3
  [arXiv:hep-ph/9710401];

  K.~Funakubo, A.~Kakuto, S.~Otsuki and F.~Toyoda,
  ``Spontaneous CP violation at finite temperature in the MSSM,''
  Prog.\ Theor.\ Phys.\  {\bf 99} (1998) 1045
  [arXiv:hep-ph/9802276];

  A.~Riotto,
  ``The more relaxed supersymmetric electroweak baryogenesis,''
  Phys.\ Rev.\  D {\bf 58} (1998) 095009
  [arXiv:hep-ph/9803357];

  K.~Funakubo,
  ``Higgs mass, CP violation and phase transition in the MSSM,''
  Prog.\ Theor.\ Phys.\  {\bf 101} (1999) 415
  [arXiv:hep-ph/9809517];

  K.~Funakubo, S.~Otsuki and F.~Toyoda,
  ``Transitional CP violation in MSSM and electroweak baryogenesis,''
  Prog.\ Theor.\ Phys.\  {\bf 102} (1999) 389
  [arXiv:hep-ph/9903276];

  J.~M.~Cline, M.~Joyce and K.~Kainulainen,
  ``Supersymmetric electroweak baryogenesis,''
  JHEP {\bf 0007} (2000) 018
  [arXiv:hep-ph/0006119];

  M.~S.~Carena, J.~M.~Moreno, M.~Quiros, M.~Seco and C.~E.~M.~Wagner,
  ``Supersymmetric CP-violating currents and electroweak baryogenesis,''
  Nucl.\ Phys.\  B {\bf 599} (2001) 158
  [arXiv:hep-ph/0011055];

  M.~S.~Carena, M.~Quiros, M.~Seco and C.~E.~M.~Wagner,
  ``Improved results in supersymmetric electroweak baryogenesis,''
  Nucl.\ Phys.\  B {\bf 650} (2003) 24
  [arXiv:hep-ph/0208043];

  T.~Prokopec, M.~G.~Schmidt and S.~Weinstock,
  ``Transport equations for chiral fermions to order h-bar and electroweak
  baryogenesis,''
  Annals Phys.\  {\bf 314} (2004) 208
  [arXiv:hep-ph/0312110];

  T.~Prokopec, M.~G.~Schmidt and S.~Weinstock,
  ``Transport equations for chiral fermions to order h-bar and electroweak
  baryogenesis. II,''
  Annals Phys.\  {\bf 314} (2004) 267
  [arXiv:hep-ph/0406140];

  T.~Konstandin, T.~Prokopec and M.~G.~Schmidt,
  ``Kinetic description of fermion flavor mixing and CP-violating sources  for
  baryogenesis,''
  Nucl.\ Phys.\  B {\bf 716} (2005) 373
  [arXiv:hep-ph/0410135];

  C.~Lee, V.~Cirigliano and M.~J.~Ramsey-Musolf,
  ``Resonant relaxation in electroweak baryogenesis,''
  Phys.\ Rev.\  D {\bf 71} (2005) 075010
  [arXiv:hep-ph/0412354];

  V.~Cirigliano, M.~J.~Ramsey-Musolf, S.~Tulin and C.~Lee,
  ``Yukawa and tri-scalar processes in electroweak baryogenesis,''
  Phys.\ Rev.\  D {\bf 73} (2006) 115009
  [arXiv:hep-ph/0603058];

  T.~Konstandin, T.~Prokopec, M.~G.~Schmidt and M.~Seco,
  ``MSSM electroweak baryogenesis and flavour mixing in transport  equations,''
  Nucl.\ Phys.\  B {\bf 738} (2006) 1
  [arXiv:hep-ph/0505103];

  V.~Cirigliano, S.~Profumo and M.~J.~Ramsey-Musolf,
  ``Baryogenesis, electric dipole moments and dark matter in the MSSM,''
  JHEP {\bf 0607} (2006) 002
  [arXiv:hep-ph/0603246];

  D.~J.~H.~Chung, B.~Garbrecht, M.~J.~Ramsey-Musolf and S.~Tulin,
  ``Yukawa Interactions and Supersymmetric Electroweak Baryogenesis,''
  Phys.\ Rev.\ Lett.\  {\bf 102} (2009) 061301
  [arXiv:0808.1144 [hep-ph]].
  
\bibitem{Carena:2008vj}
  M.~Carena, G.~Nardini, M.~Quiros and C.~E.~M.~Wagner,
  ``The Baryogenesis Window in the MSSM,''
  Nucl.\ Phys.\  B {\bf 812} (2009) 243
  [arXiv:0809.3760 [hep-ph]].

\bibitem{Funakubo:2002yb}
  K.~Funakubo, S.~Tao and F.~Toyoda,
  ``CP violation in the Higgs sector and phase transition in the MSSM,''
  Prog.\ Theor.\ Phys.\  {\bf 109} (2003) 415
  [arXiv:hep-ph/0211238].
%

\bibitem{NMSSM}
  U.~Ellwanger, M.~Rausch de Traubenberg and C.~A.~Savoy,
  ``Particle spectrum in supersymmetric models with a gauge singlet,''
  Phys.\ Lett.\  B {\bf 315} (1993) 331
  [arXiv:hep-ph/9307322];

  T.~Elliott, S.~F.~King and P.~L.~White,
  ``Radiative corrections to Higgs boson masses in the next-to-minimal
  supersymmetric Standard Model,''
  Phys.\ Rev.\  D {\bf 49} (1994) 2435
  [arXiv:hep-ph/9308309];

  T.~Moroi and Y.~Okada,
  ``Upper bound of the lightest neutral Higgs mass in extended supersymmetric
  Standard Models,''
  Phys.\ Lett.\  B {\bf 295} (1992) 73;

  J.~i.~Kamoshita, Y.~Okada and M.~Tanaka,
  ``Neutral scalar Higgs masses and production cross-sections in and extended
  supersymmetric Standard Model,''
  Phys.\ Lett.\  B {\bf 328} (1994) 67
  [arXiv:hep-ph/9402278];
  
  D.~J.~Miller, R.~Nevzorov and P.~M.~Zerwas,
  ``The Higgs sector of the next-to-minimal supersymmetric standard model,''
  Nucl.\ Phys.\  B {\bf 681} (2004) 3
  [arXiv:hep-ph/0304049];

  U.~Ellwanger, J.~F.~Gunion, C.~Hugonie and S.~Moretti,
  ``Towards a no-lose theorem for NMSSM Higgs discovery at the LHC,''
  arXiv:hep-ph/0305109;

  M.~Maniatis,
  ``The NMSSM reviewed,''
  arXiv:0906.0777 [hep-ph];

  U.~Ellwanger, C.~Hugonie and A.~M.~Teixeira,
  ``The Next-to-Minimal Supersymmetric Standard Model,''
  arXiv:0910.1785 [hep-ph].

\bibitem{Funakubo:2004ka}
  K.~Funakubo and S.~Tao,
  ``The Higgs sector in the next-to-MSSM,''
  Prog.\ Theor.\ Phys.\  {\bf 113} (2005) 821
  [arXiv:hep-ph/0409294].

\bibitem{nMSSM}
  C.~Panagiotakopoulos and K.~Tamvakis,
  ``Stabilized NMSSM without domain walls,''
  Phys.\ Lett.\  B {\bf 446} (1999) 224
  [arXiv:hep-ph/9809475];

  C.~Panagiotakopoulos and K.~Tamvakis,
  ``New minimal extension of MSSM,''
  Phys.\ Lett.\  B {\bf 469} (1999) 145
  [arXiv:hep-ph/9908351];

  C.~Panagiotakopoulos and A.~Pilaftsis,
  ``Higgs scalars in the minimal non-minimal supersymmetric standard model,''
  Phys.\ Rev.\  D {\bf 63} (2001) 055003
  [arXiv:hep-ph/0008268];

  A.~Dedes, C.~Hugonie, S.~Moretti and K.~Tamvakis,
  ``Phenomenology of a new minimal supersymmetric extension of the standard
  model,''
  Phys.\ Rev.\  D {\bf 63} (2001) 055009
  [arXiv:hep-ph/0009125];

  C.~Balazs, M.~S.~Carena, A.~Freitas and C.~E.~M.~Wagner,
  ``Phenomenology of the nMSSM from colliders to cosmology,''
  JHEP {\bf 0706} (2007) 066
  [arXiv:0705.0431 [hep-ph]].

\bibitem{UMSSM}
  D.~Suematsu and Y.~Yamagishi,
  ``Radiative symmetry breaking in a supersymmetric model with an extra U(1),''
  Int.\ J.\ Mod.\ Phys.\  A {\bf 10} (1995) 4521
  [arXiv:hep-ph/9411239];

  D.~Suematsu,
  ``Vacuum structure of the mu-problem solvable extra U(1) models,''
  Phys.\ Rev.\  D {\bf 59} (1999) 055017
  [arXiv:hep-ph/9808409];

  Y.~Daikoku and D.~Suematsu,
  ``Mass bound of the lightest neutral Higgs scalar in the extra U(1)
  models,''
  Phys.\ Rev.\  D {\bf 62} (2000) 095006
  [arXiv:hep-ph/0003205].

\bibitem{Cvetic:1997ky}
  M.~Cvetic, D.~A.~Demir, J.~R.~Espinosa, L.~L.~Everett and P.~Langacker,
  ``Electroweak breaking and the mu problem in supergravity models with an
  additional U(1),''
  Phys.\ Rev.\  D {\bf 56} (1997) 2861
  [Erratum-ibid.\  D {\bf 58} (1998) 119905]
  [arXiv:hep-ph/9703317].

\bibitem{UMSSM2}
  D.~A.~Demir, G.~L.~Kane and T.~T.~Wang,
  ``The minimal U(1)' extension of the MSSM,''
  Phys.\ Rev.\  D {\bf 72} (2005) 015012
  [arXiv:hep-ph/0503290];

  D.~A.~Demir, L.~Solmaz and S.~Solmaz,
  ``LEP indications for two light Higgs bosons and U(1)' model,''
  Phys.\ Rev.\  D {\bf 73} (2006) 016001
  [arXiv:hep-ph/0512134].

\bibitem{Erler:2002pr}
  J.~Erler, P.~Langacker and T.~j.~Li,
  ``The Z - Z' mass hierarchy in a supersymmetric model with a secluded
  U(1)'-breaking sector,''
  Phys.\ Rev.\  D {\bf 66}, 015002 (2002)
  [arXiv:hep-ph/0205001].

\bibitem{Han:2004yd}
  T.~Han, P.~Langacker and B.~McElrath,
  ``The Higgs sector in a U(1)' extension of the MSSM,''
  Phys.\ Rev.\  D {\bf 70} (2004) 115006
  [arXiv:hep-ph/0405244].

\bibitem{Kang:2004pp}
  J.~Kang, P.~Langacker, T.~j.~Li and T.~Liu,
  ``Electroweak baryogenesis in a supersymmetric U(1) -prime model,''
  Phys.\ Rev.\ Lett.\  {\bf 94} (2005) 061801
  [arXiv:hep-ph/0402086].

\bibitem{Chiang:2008ud}
  C.~W.~Chiang and E.~Senaha,
  ``CP violation in the secluded U(1)'-extended MSSM,''
  JHEP {\bf 0806} (2008) 019
  [arXiv:0804.1719 [hep-ph]].

\bibitem{Kang:2009rd}
  J.~Kang, P.~Langacker, T.~Li and T.~Liu,
  ``Electroweak Baryogenesis, CDM and Anomaly-free Supersymmetric U(1)-prime
  Models,''
  arXiv:0911.2939 [hep-ph].
  
\bibitem{Funakubo:2005pu}
  K.~Funakubo, S.~Tao and F.~Toyoda,
  ``Phase transitions in the NMSSM,''
  Prog.\ Theor.\ Phys.\  {\bf 114} (2005) 369
  [arXiv:hep-ph/0501052].

\bibitem{Menon:2004wv}
  A.~Menon, D.~E.~Morrissey and C.~E.~M.~Wagner,
  ``Electroweak baryogenesis and dark matter in the nMSSM,''
  Phys.\ Rev.\  D {\bf 70} (2004) 035005
  [arXiv:hep-ph/0404184];

\bibitem{Huber:2006wf}
  S.~J.~Huber, T.~Konstandin, T.~Prokopec and M.~G.~Schmidt,
  ``Electroweak Phase Transition and Baryogenesis in the nMSSM,''
  Nucl.\ Phys.\  B {\bf 757} (2006) 172
  [arXiv:hep-ph/0606298].
  
  \bibitem{EWPT_UMSSM}
  S.~W.~Ham, E.~J.~Yoo and S.~K.~OH,
  ``Electroweak phase transitions in the MSSM with an extra $U(1)'$,''
  Phys.\ Rev.\  D {\bf 76} (2007) 075011
  [arXiv:0704.0328 [hep-ph]];

  S.~W.~Ham and S.~K.~OH,
  ``Electroweak phase transition in MSSM with $U(1)'$ in explicit CP violation
  scenario,''
  Phys.\ Rev.\  D {\bf 76} (2007) 095018
  [arXiv:0708.1785 [hep-ph]].

  
  \bibitem{Erler:2000wu}
  J.~Erler,
  ``Chiral models of weak scale supersymmetry,''
  Nucl.\ Phys.\  B {\bf 586} (2000) 73
  [arXiv:hep-ph/0006051].

\bibitem{Kang:2004bz}
  J.~Kang and P.~Langacker,
  ``Z' discovery limits for supersymmetric E(6) models,''
  Phys.\ Rev.\  D {\bf 71} (2005) 035014
  [arXiv:hep-ph/0412190].


\bibitem{Coleman:1973jx}
  S.~R.~Coleman and E.~J.~Weinberg,
  ``Radiative Corrections As The Origin Of Spontaneous Symmetry Breaking,''
  Phys.\ Rev.\  D {\bf 7} (1973) 1888.

\bibitem{Zprime_search}
  J.~Erler, P.~Langacker, S.~Munir and E.~R.~Pena,
  ``Improved Constraints on Z' Bosons from Electroweak Precision Data,''
  JHEP {\bf 0908} (2009) 017
  [arXiv:0906.2435 [hep-ph]];

  J.~Erler, P.~Langacker, S.~Munir and E.~Rojas,
  ``Constraints on the mass and mixing of $Z'$ bosons,''
  arXiv:0910.0269 [hep-ph].


\bibitem{Funakubo:2009eg}
  K.~Funakubo and E.~Senaha,
  ``Electroweak phase transition, critical bubbles and sphaleron decoupling
  condition in the MSSM,''
  Phys.\ Rev.\  D {\bf 79} (2009) 115024
  [arXiv:0905.2022 [hep-ph]].

\bibitem{Funakubo:2005bu}
  K.~Funakubo, A.~Kakuto, S.~Tao and F.~Toyoda,
  ``Sphalerons in the NMSSM,''
  Prog.\ Theor.\ Phys.\  {\bf 114} (2006) 1069
  [arXiv:hep-ph/0506156].

\bibitem{Dolan:1973qd}
  L.~Dolan and R.~Jackiw,
  ``Symmetry Behavior At Finite Temperature,''
  Phys.\ Rev.\  D {\bf 9} (1974) 3320.

\bibitem{Anderson:1991zb}
  G.~W.~Anderson and L.~J.~Hall,
  ``The Electroweak Phase Transition And Baryogenesis,''
  Phys.\ Rev.\  D {\bf 45} (1992) 2685.

\bibitem{SM_cutoff}
  C.~Grojean, G.~Servant and J.~D.~Wells,
 ``First-order electroweak phase transition in the standard model with a  low
  cutoff,''
  Phys.\ Rev.\  D {\bf 71} (2005) 036001
  [arXiv:hep-ph/0407019];

  D.~Bodeker, L.~Fromme, S.~J.~Huber and M.~Seniuch,
  ``The baryon asymmetry in the standard model with a low cut-off,''
  JHEP {\bf 0502} (2005) 026
  [arXiv:hep-ph/0412366].

\bibitem{Pietroni:1992in}
  M.~Pietroni,
  ``The Electroweak phase transition in a nonminimal supersymmetric model,''
  Nucl.\ Phys.\  B {\bf 402} (1993) 27
  [arXiv:hep-ph/9207227].

\bibitem{Huber:2000mg}
  S.~J.~Huber and M.~G.~Schmidt,
  ``Electroweak baryogenesis: Concrete in a SUSY model with a gauge  singlet,''
  Nucl.\ Phys.\  B {\bf 606} (2001) 183
  [arXiv:hep-ph/0003122].

\end{thebibliography}
\end{document}